\documentclass[lettersize,journal]{IEEEtran}
\usepackage{amsmath,amsfonts}
\usepackage{algorithmic}
\usepackage{algorithm}
\usepackage{array}
\usepackage[caption=false,font=normalsize,labelfont=sf,textfont=sf]{subfig}
\usepackage{textcomp}
\usepackage{stfloats}
\usepackage{url}
\usepackage{verbatim}
\usepackage{graphicx}
\usepackage{booktabs}
\usepackage{cite}
\usepackage{multirow}
\begin{document}

\title{Reducing End-to-End Latency of Cause-Effect Chains with Shared Cache Analysis

}

\author{ Yixuan Zhu, Yinkang Gao, Bo Zhang, Xiaohang Gong, Binze Jiang, Lei Gong~\IEEEmembership{Member,~IEEE,} Wenqi Lou, Teng Wang, Chao Wang~\IEEEmembership{Senior Member,~IEEE,} Xi Li, Xuehai Zhou
\thanks{Yixuan Zhu, Yinkang Gao, Bo Zhang, Xiaohang Gong, Binze Jiang, Lei Gong, Wenqi Lou, Chao Wang, Xi Li, and Xuehai Zhou are with the University of Science and Technology of China, Hefei, Anhui 230026, China, and also with Suzhou Institute for Advanced Research, University of Science and Technology of China, Suzhou 215123, China 
(e-mail: \{zhuyixuan, gaoyinkang, sazb, gxh2018, jiangbinze\}@mail.ustc.edu.cn; \{leigong0203, louwenqi, wangt635, cswang, llxx, xhzhou\}@ustc.edu.cn).}
}

\markboth{IEEE Transactions on computers}%
{Shell \MakeLowercase{\textit{et al.}}: A Sample Article Using IEEEtran.cls for IEEE Journals}


\maketitle

\begin{abstract}
Cause-effect chains, as a widely used modeling method in real-time embedded systems, are extensively applied in various safety-critical domains. End-to-end latency, as a key real-time attribute of cause-effect chains, is crucial in many applications. But the analysis of end-to-end latency for cause-effect chains on multicore platforms with shared caches still presents an unresolved issue. Traditional methods typically assume that the worst-case execution time (WCET) of each task in the cause-effect chain is known. However, in the absence of scheduling information, these methods often assume that all shared cache accesses result in misses, leading to an overestimation of WCET and, consequently, affecting the accuracy of end-to-end latency. However, effectively integrating scheduling information into the WCET analysis process of the chains may introduce two challenges: first, how to leverage the structural characteristics of the chains to optimize shared cache analysis, and second, how to improve analysis accuracy while avoiding state space explosion.

To address these issues, this paper proposes a novel end-to-end latency analysis framework designed for multi-chain systems on multicore platforms with shared caches. This framework extracts scheduling information and structural characteristics of cause-effect chains, constructing fine-grained and scalable inter-core memory access contexts at the basic block level for time-sensitive shared cache analysis. This results in more accurate WCET (TSC-WCET) estimates, which are then used to derive the end-to-end latency. Finally, we conduct experiments on dual-core and quad-core systems with various cache configurations, which show that under certain settings, the average maximum end-to-end latency of cause-effect chains is reduced by up to 34\% and 26\%.
\end{abstract}

\begin{IEEEkeywords}
Shared cache, cause-effect chain, end to end latency, multicore System.
\end{IEEEkeywords}

\section{Introduction}
%
%
%
%
\IEEEPARstart{W}{ith} the rapid development of domains such as automotive electronics and industrial control, real-time embedded systems~\cite{Apollo} evolving toward higher functionality and greater integration. These systems typically involve end-to-end processes composed of multiple cooperating tasks, including sensor data acquisition, information processing, decision making, and actuator control~\cite{li2024data}. To accurately capture the data dependencies and execution order among these tasks, and to effectively evaluate the system’s real-time characteristics, the modeling and analysis approach known as cause-effect chains has been proposed and widely adopted.

In the cause-effect chain model, system functionality is abstracted as a directed acyclic linear graph, where each node represents a task and edges denote causal dependencies between tasks. Based on the triggering mechanism of non-initial tasks, cause-effect chains can be classified into time-triggered (TT) chains and event-triggered (ET) chains. In both types, the first task is a periodically triggered sampling task. In TT chains, subsequent tasks are independently activated using the same predefined period; in contrast, in ET chains, each subsequent task is triggered immediately upon the completion of its direct predecessor. The real-time property of a cause-effect chain is typically characterized by its end-to-end latency\footnote{In fact, the property described here corresponds to only one aspect of end-to-end latency—data age—and does not cover the other important attribute, reaction time. However, these two metrics are functionally interchangeable and can be converted into one another~\cite{gunzel2023equivalence}.}, defined as the time elapsed from the start of the first task to the completion of the last task in the chain.

Traditional end-to-end latency analysis methods~\cite{guenzel2024end, davare2007period,durr2019end,zhu2013optimization,gunzel2021timing} are generally built upon a fundamental assumption: the WCET~\cite{LiWCET} of each task in a cause-effect chain is known in advance. However, on multicore platforms with shared cache, such methods—lacking concrete scheduling information—must still ensure analysis safety. As a result, they often adopt extremely conservative assumptions for shared cache behavior (e.g., assuming all accesses to the shared cache are misses), which significantly inflates the estimated WCET. This pessimism further propagates to the end-to-end latency, causing the analysis results to deviate from actual system behavior severely. We refer to this issue as: \textbf{ Latency pessimism induced by schedule-independent WCET assumptions}. 

To address the aforementioned issues, it is imperative to effectively incorporate scheduling information into the WCET analysis. However, achieving this goal presents \textbf{Two Key Challenges}: \textbf{First, how to effectively leverage the characteristics of cause-effect chains to optimize shared cache analysis}. The core of shared cache analysis lies in modeling inter-core memory access interference, which heavily depends on accurately estimating the timing of memory accesses on other cores—constructing an inter-core context model. Cause-effect chains inherently exhibit a linear topology, and the activation times of certain tasks are deterministically known, offering a feasible basis for more precise temporal context modeling. However, existing methods have yet to systematically integrate such information, resulting in analysis outcomes that remain overly conservative. \textbf{Second, how to enhance shared cache analysis precision while avoiding state space explosion}~\cite{maiza2019survey}. Due to the effects of scheduling, program loops, and branches, the actual occurrence time of memory accesses is highly non-deterministic, significantly increasing the complexity of context modeling. To avoid state space explosion, existing approaches typically adopt coarse-grained time modeling~\cite{hardy2009using,fischer2023analysis,nagar2014precise,liang2012timing,zhang2022precise} strategies, treating the entire program execution interval as the potential window for all memory accesses. While this approximation improves scalability, it overlooks fine-grained temporal characteristics within programs. It tends to misclassify a large number of infeasible memory access overlaps as interference, thereby further exacerbating the pessimism of the analysis results.

To this end, this paper proposes a novel end-to-end latency analysis framework targeting multi-chains partitioned scheduling on shared cache multi-core platforms. The core idea is to effectively exploit the scheduling information and characteristics inherent in cause-effect chains (Challenge 1), and to construct fine-grained inter-core contexts at the basic block level for time-sensitive shared cache analysis with relative and approximate timing models (Challenge 2), thereby enabling a more accurate WCET (TSC-WCET) and ultimately deriving a tight end-to-end latency. Specifically, the framework begins by determining the task execution sequence on each core based on scheduling information. Then, leveraging control flow graphs (CFGs) and architectural characteristics, it applies a path-based analysis to derive the possible execution time intervals of basic blocks or virtual nodes (child loops transformed), relative to its current loop. To avoid state space explosion, a relative time modeling mechanism is introduced, which uses the CFG, task execution order, and release times to approximate a basic block’s execution window with respect to the system start time, serving as the context. In the interference identification phase, the framework introduces a hierarchical strategy, which incrementally determines potential temporal overlaps with the target memory access across three levels: task instances, loops, and basic blocks, enabling fine-grained and scalable. Furthermore, the framework incorporates mutually exclusive basic block information both within and across programs to optimize interference estimation, eliminating interferences that cannot occur concurrently, thus improving the accuracy of the analysis. Then it is integrated with pipeline analysis to estimate the execution costs of basic blocks, which are then used in an Integer Linear Programming model to compute the TSC-WCET for each task. Finally, based on the obtained WCETs and scheduling information, we derive the end-to-end latency for chains. Experiment results show that our method consistently outperforms existing approaches, reducing average maximum end-to-end latency by up to 34\% on dual-core and 26\% on quad-core systems under certain settings. The main contributions of this paper are as follows:
\begin{itemize}
    \item A novel end-to-end latency analysis framework for multi-core systems with shared caches, which effectively integrates scheduling information into WCET analysis, particularly the shared cache analysis, thereby eliminating the pessimistic assumptions inherent in prior approaches.
    \item A relative and approximate time-based memory access context modeling method at the basic block level is designed, which balances analysis precision and scalability, laying the foundation for accurate shared cache interference analysis.
    \item A novel interference reduction mechanism is introduced, leveraging both intra- and inter-program exclusive execution basic blocks to reduce overestimation of interference and enhance the tightness of the analysis.
    \item A formal safety proof of the proposed method is provided, ensuring that the end-to-end latency remains safe under inter-core interference modeling.

\end{itemize}

\section{Related Work}
End-to-end latency analysis of cause-effect chains has been extensively studied in real-time systems. For ET chains, Chakraborty et al.~\cite{chakraborty2003general} proposed a framework called Real-Time Calculus, which models generic event streams to analyze various timing properties, including end-to-end latency. Subsequently, Henia et al.~\cite{henia2005system} introduced an industrial analysis method and toolchain, SymTA/S, for analyzing timing behavior in embedded systems, which is particularly effective for ET models. For TT chains, Davare et al.~\cite{davare2007period} analyzed the end-to-end latency of TT chains and proposed an analytical bound. They also studied how to assign task periods to optimize end-to-end latency. In 2019, Dürr et al.~\cite{durr2019end} extended this bound to sporadic task models and proposed a tighter upper bound. Later, Dürr et al.~\cite{gunzel2021timing} analyzed the end-to-end latency of interconnected cause-effect chains across asynchronous processors under the assumption of fixed task execution times. In terms of system applications, Teper~\cite{teper2022end} proposed an end-to-end analysis approach for systems such as ROS2~\cite{ROS2} that involve both TT and ET chains. Regarding task triggering modes, Tang~\cite{tang2023reaction} proposed a novel triggering model, ETDR, which combines ET with data refreshing to address the limitations of traditional TT and ET models. ETDR effectively handles data loss and task skipping while optimizing reaction time.

Although the above methods support the analysis of various types of cause-effect chains and scheduling models, they typically assume that the WCET of each task is known in advance, and then derive the chain's end-to-end latency based on scheduling information. However, in multi-core systems with shared caches, the WCET of a task is affected by tasks executing on other cores. Due to the lack of accurate scheduling information during WCET analysis, these approaches cannot effectively model inter-core tasks' interference, which may lead to overly pessimistic analysis results.

For cache analysis in single-core systems, researchers~\cite{alt1996cache, ferdinand1999efficient} have widely adopted abstract interpretation~\cite{cousot1977abstract} techniques to model cache behavior and improve the accuracy of WCET analysis. Related work~\cite{hardy2008wcet} has also been extended to multi-level and set-associative caches. In addition, some approaches support cache-related preemption delay analysis for preemptive scheduling on single-core systems~\cite{altmeyer2009new,staschulat2007scalable}. In multi-core systems, the interference caused by concurrently executing programs on other cores makes shared cache analysis particularly challenging. Yan et al.~\cite{yan2008wcet} first introduced a shared cache analysis for WCET in 2008, assuming that all tasks running on other cores cause interference, which leads to high pessimism. Li and Liang~\cite{liang2012timing} improved the WCRT analysis by considering task lifetimes to exclude infeasible interfering tasks, and extended their method to set-associative caches. Nagar~\cite{nagar2014precise, nagar2016fast} proposed an ILP-based shared cache interference analysis that statically determines the maximum number of interfering accesses from competing tasks and computes the worst-case interference placement on the CFG to derive the maximum execution time increase. This method provides more accurate WCET analysis. Zhang et al.~\cite{zhang2022precise} proposed an improved shared cache interference analysis by introducing a happens-before partial order on memory accesses in the CFG to eliminate infeasible interference combinations, thereby improving analysis accuracy. Fischer~\cite{fischer2023analysis} proposed a method that leverages arrival curves to model interference and introduced a novel dataflow-based classification approach for cache hit/miss behavior in shared caches.

Although the above shared cache analysis methods improve the accuracy of WCET or WCRT estimation by optimizing interference analysis from various perspectives, they fail to effectively exploit the structural characteristics of cause-effect chains to further enhance analysis precision.

\begin{figure}
    \centering
    \includegraphics[width=1\linewidth]{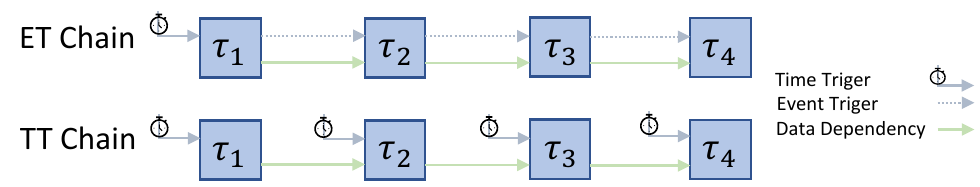}
    \caption{Event-triggered Chain and Time-triggered Chain}
    \label{fig:EE_TT_Chain}
    \vspace{-10pt}
\end{figure}

\section{Background \& Motivation}
\subsection{Background}
\subsubsection{Hardware Architecture and System Model} 
\
\newline
\indent
The target platform considered in this work is a typical multi-core architecture, where each core is equipped with a private L1 cache, and all cores share a L2 cache. The L2 cache is set-associative and employs the Least Recently Used (LRU) replacement policy. We focus on instruction caches, as the memory layout of program code is known at compile time. Nevertheless, the proposed analysis approach is general and can be applied to data and unified caches alike.

This work adopts the cause-effect chain as the fundamental task model. A cause-effect chain is defined as an ordered sequence of tasks, denoted by $\Gamma = \langle \tau_1, \tau_2, \dots, \tau_n \rangle$, where each task $\tau_i$ has a data dependency on its immediate predecessor $\tau_{i-1}$. Typically, $\tau_1$ is a periodically triggered sampling task that serves as the starting point of the entire chain. In TT chains, the subsequent tasks are also triggered periodically, whereas in ET chains, each task is activated immediately after the completion of its direct predecessor, as shown in Fig.~\ref{fig:EE_TT_Chain}.

To simplify the complexity of timing analysis, this paper focuses on ET chains and a special form of TT chains—back-to-back TT chains~\cite{kohler2023robust}. In such chains, all tasks have a same activation period $T_{\Gamma}$, and task executions are aligned through carefully configured time offsets (with task $i$ assigned an offset $off_i$) such that each task is triggered immediately after the completion of its predecessor, which avoids execution overlap between tasks and enhances system predictability.

The system consists of multiple independent cause-effect chains, each statically assigned to a processing core. We assume that each core executes only one cause-effect chain to simplify the analysis and eliminate uncertainties arising from scheduling interference or mixed triggering mechanisms among multiple chains a core. Each chain is assigned a static priority, which is inherited by all tasks within the chain and used for local priority-based scheduling on the core. In fact, for more general scenarios—where each core executes multiple cause-effect chains of the same triggering type and with the same release period—these chains can be merged into a single, longer cause-effect chain. This transformation enables the application of the proposed analysis method to such cases. Therefore, the assumptions made in this work not only ensure analytical simplicity but also retain a degree of extensibility.

Given a specific job instance of a task chain $\Gamma $, let $s_1^{(j)}$ denote the start time of task $\tau_1^{(j)}$ and $f_n^{(j)}$ denote the finish time of the last task $\tau_n^{(j)}$ in $\Gamma$. Then, the end-to-end latency (EL) of $\Gamma$ is defined as:
$$
L_{\Gamma}^{(j)} = f_n^{(j)} - s_1^{(j)}
$$

The maximum end-to-end latency (MEL) is defined as:
$$
L_{\Gamma}^{\max} = \sup_j \left( f_n^{(j)} - s_1^{(j)} \right)
$$

\begin{table}[!t]
\caption{Cache Behavior Characteristics}
\label{tab:cache_behavior}
\centering
\begin{tabular}{c c c}
\toprule
\textbf{Classification} & \textbf{Behavioral Characteristics} & \textbf{Time Cost} \\
\midrule
Always Hit (AH)      & Every access must hit           & access cache   \\
Always Miss (AM)     & Every access must not hit       & access memory  \\
Persistent (PS)      & First is hit, otherwise miss    & access cache   \\
Not Classified (NC)  & Not categorized as first three  & both consider  \\
\bottomrule
\end{tabular}
\label{label:cacheBehavior}
\vspace{-10pt}
\end{table}

\subsubsection{Shared Cache Analysis}
\
\newline
\indent
Shared cache analysis determines the Cache Hit/Miss Classification (CHMC)~\cite{cullmann2013cache}, which characterizes the execution latency of memory access operations. As shown in Table~\ref{label:cacheBehavior}, each CHMC category is associated with specific access characteristics and corresponding timing costs. It is important to note that for memory accesses classified as NC, both cache hit and miss latencies must be considered in subsequent analysis to ensure the safety of the results. In general, shared cache analysis typically consists of the following two steps~\cite{yan2008wcet}.

In the first step, the influence of inter-core shared resources is ignored by assuming that the current core has exclusive access to the shared cache. Under this assumption, the CHMC is directly obtained using single-core multi-level cache analysis techniques based on abstract interpretation~\cite{theiling1998combining}. For memory accesses that hit in the shared cache, it is also necessary to compute the corresponding maximum age $age(m)$ under the LRU replacement policy.

In the second step, the effect of shared resources is taken into account, and the previously computed CHMC is refined based on inter-core context and interference. If memory accesses from multiple cores are mapped to the same cache set, data stored by the current core in that set may be evicted. This process affects only memory accesses with CHMC classified as AH or PS, and if a given condition (denoted as Eqn.~\ref{eqn:changeCHMC}) is satisfied, the CHMC of access $m$ should be updated to NC. Here, $N$ denotes the ways of cache and $|{\mathcal M}_c(m)|$ represents the number of interfering memory accesses from other cores that may evict the cache block corresponding to $m$.
\begin{equation}
\setlength\abovedisplayskip{3pt}
\setlength\belowdisplayskip{3pt}
N - age(m) < |{\mathcal M}_c(m)|
\label{eqn:changeCHMC}
\end{equation}

In Eqn.~\ref{eqn:changeCHMC}, $N$ is known and the maximum age $age(m)$ is obtained from the first analysis step. The key challenge lies in accurately estimating the interference count $|{\mathcal M}_c(m)|$. Underestimating this value may compromise the safety of subsequent timing analysis, as memory accesses that actually result in cache misses may be incorrectly classified as hits. On the other hand, overestimating the interference count introduces significant pessimism, potentially misclassifying actual hits as misses and thereby reducing the tightness of the analysis.

This overestimation issue also arises in the maximum end-to-end latency analysis of cause-effect chains. In earlier approaches, the WCET is typically assumed to be known, but such WCET are often derived from conservative shared cache analyses that lack scheduling information, leading to overly pessimistic results. Therefore, accurately estimating the interference count while ensuring analysis safety remains a fundamental challenge in shared cache interference analysis.

\subsection{Motivation}

\begin{figure*}
    \centering
    \includegraphics[width=0.95\linewidth]{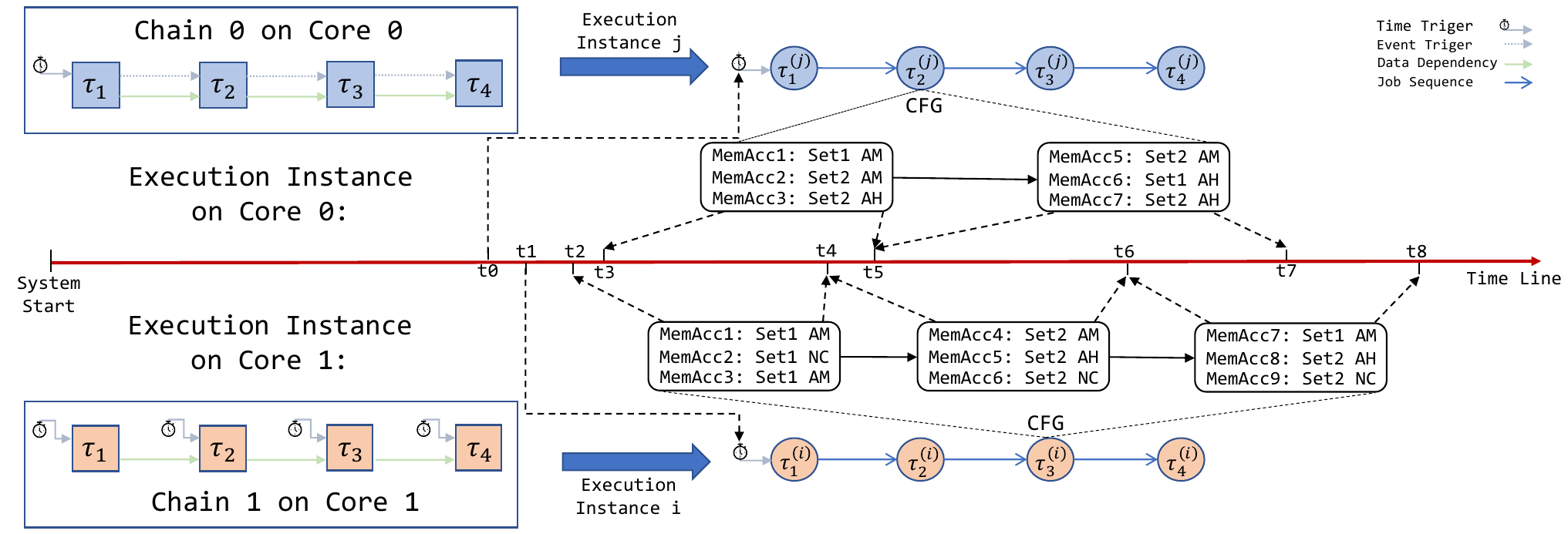}
    \caption{Information for access shared cache on CFG for two jobs from the corresponding chains and map their execution time to timeline}
    \label{fig:motivation}
    \vspace{-10pt}
\end{figure*}

This paper explores the integration of scheduling information into shared cache analysis by leveraging the structural characteristics of cause-effect chains, resulting in a time-sensitive cache analysis method (TSC). This enables the derivation of TSC-aware Worst-Case Execution Time (TSC-WCET), which is then used to compute a tighter  end-to-end latency. For general systems—such as multi-core DAG~\cite{sun2023real} systems under dynamic scheduling or multi-rate cause-effect chain systems under global scheduling~\cite{gunzel2023compositional}—this approach would suffer from severe state space explosion: the combination of microarchitectural states with scheduling-level states leads to exponential growth in the analysis space. However, for the specific class of systems targeted in this work—namely, independent cause-effect chains operating under fixed-priority partitioned scheduling—the analysis state space is significantly reduced. This makes it feasible to perform fine-grained and high-precision time-sensitive cache interference modeling.

Under this system model, the task scheduling sequence can be statically determined and is unique, which facilitates the analysis of TSC-WCET. The partitioned scheduling strategy stipulates that each cause-effect chain is statically assigned to a dedicated core, and only one chain is executed per core. For the extended scenarios discussed earlier, in which multiple cause-effect chains of the same type and period are executed on the same core, these chains—due to their identical sampling periods—can be concatenated into a single longer chain based on their priorities. In the case of TT chains, back-to-back execution can be enforced by properly configuring the offset of tasks, thereby enabling their merger into a chain as well. Through these chain-merging strategies, only one deterministic job chain per core needs to be analyzed. As illustrated in Fig.~\ref{fig:motivation}, the instantiated job chains on cores 0 and 1 are both uniquely defined. This significantly reduces the state space for interference analysis, thereby creating favorable conditions for subsequent cache and end-to-end latency analysis.

Cause-effect chains inherently carry timing information, which provides favorable conditions for fine-grained shared cache analysis. In such an analysis, computing the interference requires determining whether a given memory access may interfere with memory accesses from other cores. This process involves checking not only spatial overlap—whether the accesses map to the same cache set—but also temporal overlap—whether they occur within overlapping time intervals. However, existing analysis methods often lack fine-grained timing information, making it difficult to accurately determine such temporal overlaps and thereby reducing the precision of interference classification. Fig.~\ref{fig:motivation} shows two cause-effect chains running on core 0 and core 1, respectively, with the execution times of basic blocks from specific task instances mapped onto a unified time axis. Each basic block is annotated with shared cache access information (〈access ID: accessed cache set, CHMC from the first analysis step〉), as well as its execution time relative to the system start. In traditional WCET analysis without fine-grained timing information, a memory access on core 0 (e.g., MemAcc1) is typically assumed to be potentially interfered with by all memory accesses from core 1 that target the same cache set (e.g., set 1), regardless of their actual timing. These approaches~\cite{guenzel2024end, davare2007period,durr2019end,zhu2013optimization,gunzel2021timing} clearly overestimate the interference, as many of those task instances may not overlap in time. In contrast, for cause-effect chains, the linear topology and the periodic release characteristics make it more feasible to derive the execution time windows of specific instances. This, in turn, enables more accurate identification of temporal overlaps between memory accesses, significantly improving the precision of interference analysis.

We further observe that even if two task instances overlap in time, their internal basic blocks may not necessarily execute concurrently. For example, although the MemAcc2 on core 0 and MemAcc9 on core 1 target the same cache set, their respective basic blocks do not overlap in time, and thus no actual interference occurs. However, existing shared cache analysis methods~\cite{hardy2009using,fischer2023analysis,nagar2014precise,liang2012timing,zhang2022precise} misclassify such cases as interference simply because programs P0 and P1 are concurrently executing. Moreover, a common issue in current approaches is the uniform estimation of interference intensity: all memory accesses mapped to the same cache set are assumed to suffer the same degree of interference. To address these problems, we propose leveraging the possible execution time intervals of basic blocks as fine-grained timing information for their internal memory accesses. To address the state space explosion caused by precisely enumerating all possible execution times of basic blocks under branches, loops, and other control-flow constructs, we propose constructing an approximate yet safe execution time window based on the notions of earliest start time and latest finish time.

\begin{figure*}
    \centering
    \includegraphics[width=1\linewidth]{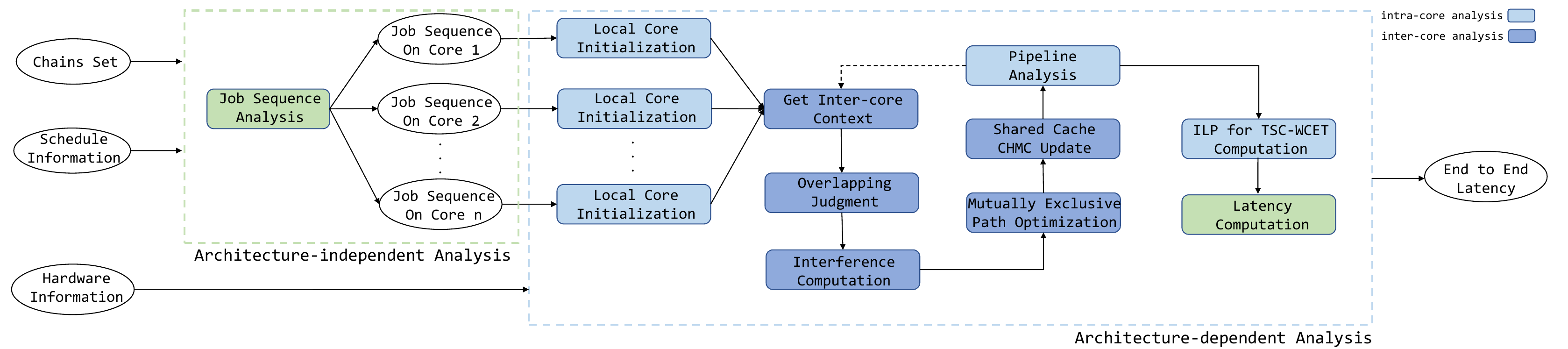}
    \caption{Overview of our analysis framework}
    \label{fig:framework}
    \vspace{-10pt}
\end{figure*}

\section{Method}

\subsection{Overall Framework }
This section first introduces our analysis framework, as illustrated in Fig.~\ref{fig:framework}. The inputs to the framework include the binary code of tasks in the set of cause-effect chains, scheduling information, and relevant hardware characteristics. The output is the maximum end-to-end latency for each cause-effect chain. The overall analysis consists of two main components: architecture-independent and architecture-dependent.

The architecture-independent analysis, represented by the green dashed box in the figure, is primarily based on the provided scheduling information, including the static mapping of cause-effect chains to processing cores, task priorities, and the chains' activation periods. From this, a periodic task execution sequence can be derived for each core, corresponding to the tasks of the cause-effect chains assigned to it.

The architecture-dependent analysis, delineated by the blue dashed box in the figure, focuses on analyzing both intra-core and inter-core shared cache interference based on the task execution sequences on each core and the underlying hardware configuration. Under a pessimistic assumption, the analysis begins by initializing each task within the task chains on each core to obtain the best-case ($BBBC()$) and worst-case ($BBWC()$) execution costs of their internal basic blocks. Leveraging these costs and the scheduling information, the potential execution time intervals of each basic block, relative to the system start time, are derived and used as the execution context for subsequent memory access analysis. In the overlap detection module, the execution contexts are utilized to determine the temporal overlap relationships between basic blocks, and the results are passed to the interference computation module to estimate the number of potential interferences. Then, using mutual exclusion path information, pairs of basic blocks that cannot execute concurrently are identified to exclude infeasible interferences, thereby improving the accuracy of interference computation. Based on the number of interferences, a refined shared cache analysis is conducted to update the CHMC, followed by a new pipeline analysis to obtain updated basic block execution costs. These updated execution costs serve as inputs for subsequent analyses, including the computation of task TSC-WCET within each job sequence using an ILP model, and the estimation of maximum end-to-end latency over the hyper-period. Ultimately, the framework outputs the latency for each cause-effect chain.

\subsection{Initialization }

\begin{figure}
    \centering
    \includegraphics[width=1\linewidth]{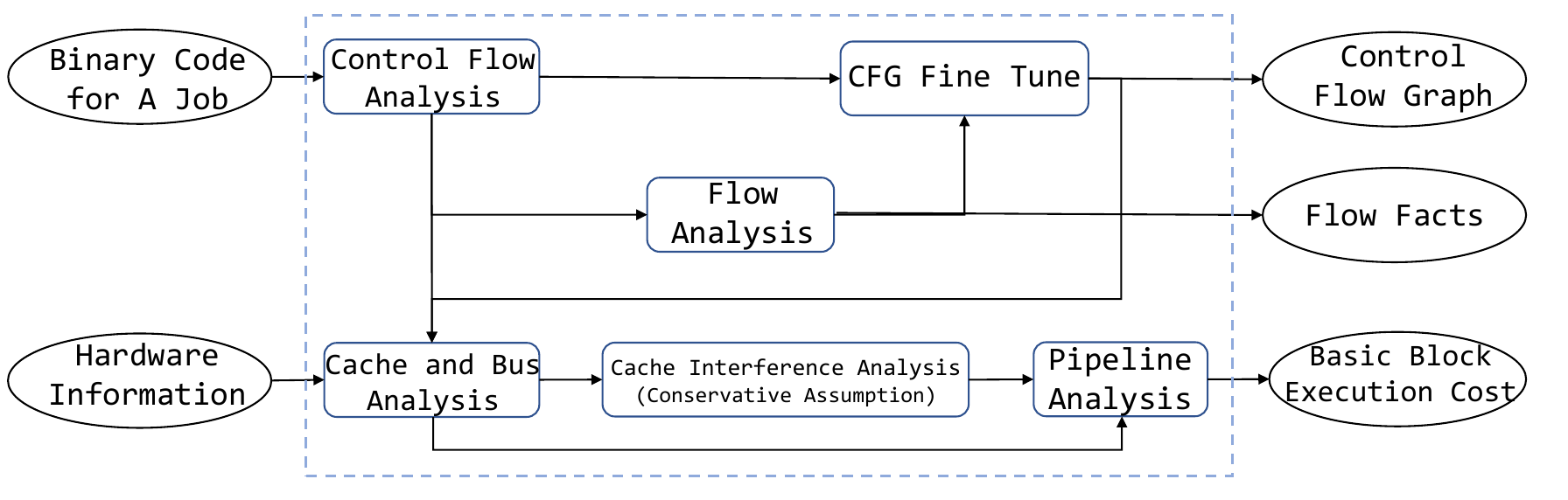}
    \caption{Initialization Module}
    \label{fig:Initial}
    \vspace{-10pt}
\end{figure}

At this stage, the binary code of the program and hardware characteristics are utilized to estimate the basic blocks' execution costs under conservative interference assumptions, while producing an optimized CFG and corresponding flow facts. All tasks undergo the same initialization procedure, as illustrated in Fig.~\ref{fig:Initial}. 

First, the control flow analysis module processes the binary code to extract the CFG and associated flow information. We leverage the module from the SWEET~\cite{lisper2014sweet} framework for flow analysis, which provides information such as loop upper bounds ($MaxBd(L_X)$), loop lower bounds ($MinBd(L_X)$), and mutually exclusive basic blocks. Flow facts are manually supplemented for cases not adequately handled by the SWEET. Subsequently, based on the upper and lower bound constraints, the CFG is refined to eliminate invalid edges introduced by VIVU~\cite{martin1998analysis} during private cache analysis.
Architecture-dependent low-level timing analysis is then performed using hardware specifications and program characteristics. This includes analysis of the cache and the bus. The CHMC for cache is then updated using the conservative assumptions about shared cache interference. Specifically, When pipeline analysis, the best-case basic block execution cost ($BBBC()$) is computed under the assumption of no interference for shared cache, i.e., all accesses to the shared cache result in hits; conversely, the worst-case cost ($BBWC()$) is computed under the assumption of infinite interference, i.e., all shared cache accesses result in misses. The pipeline analysis adopts an interval-based timing analysis approach~\cite{li2006modeling}.

\subsection{Get Inter-core Context}
\begin{figure}
    \centering
    \includegraphics[width=1\linewidth]{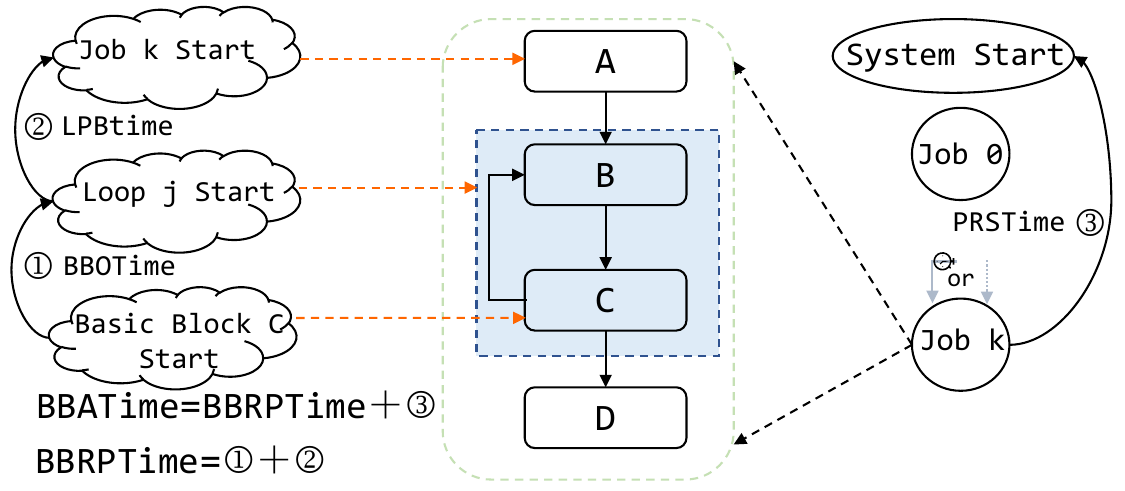}
    \caption{Derive the execution time of a basic block relative to system start through multi-level relative timing}
    \label{fig:relativeRepresent}
    \vspace{-10pt}
\end{figure}

As established earlier, obtaining the inter-core execution context is equivalent to determining the potential execution time of each basic block relative to the system start, referred to as the basic block's absolute execution time ($BBATime$). In this work, we adopt the notions of relative time and offset time to represent temporal information. As illustrated in Fig.~\ref{fig:relativeRepresent}, the $BBATime$ of a basic block can be derived by combining its execution time relative to the start of its program ($BBRPTime$) with the program's start time relative to the system ($PRSTime$).

The $BBRPTime$ is represented using the basic block’s offset time ($BBOTime$) and the loop’s base time ($LPBTime$), which enables the decomposition of the overall CFG into subgraphs for analysis, thereby simplifying the problem and enhancing parallelism. For basic blocks within a loop, $BBOTime$ refers to their execution time relative to the start of their immediate enclosing loop. For basic blocks not enclosed in any loop, $BBOTime$ refers to their execution time relative to the program start, equivalent to $BBRPTime$. The $LPBTime$ denotes the start time of a loop concerning the program start and is described by the loop’s relative time ($LPRTime$). The $LPRTime$ indicates the start time of a loop relative to the start of its immediate parent loop.

For any basic block outside of a loop, its $BBOTime$ is represented as an interval that characterizes the range of its possible execution times. Due to the influence of conditional branches, a basic block may have multiple possible start and end times. Such combinations grow exponentially with the number of branches, making explicit enumeration infeasible. To ensure scalability, we adopt an approximate representation of the offset time for a basic block $k$ using its earliest possible start time ($BBESOT(k)$) and latest possible end time ($BBLEOT(k)$), denoted as the interval $[BBESOT(k), BBLEOT(k)]$.

\begin{figure}
    \centering
    \includegraphics[width=1\linewidth]{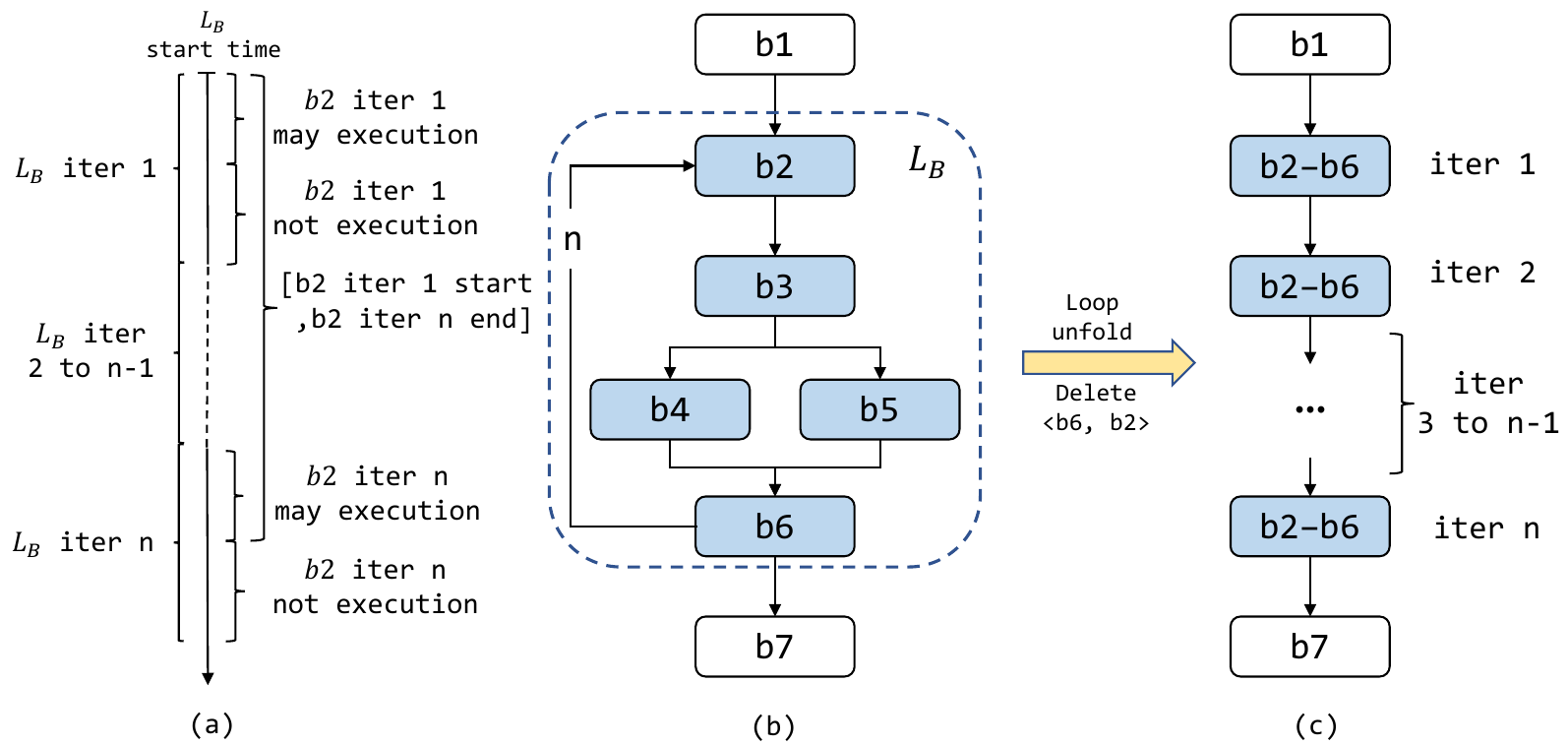}
    \caption{Execution time representation of basic blocks within a loop}
    \label{fig:innerLoopReprent}
    \vspace{-15pt}
\end{figure}

The $BBOTime$ of a basic block within a loop is represented as a sequence of intervals corresponding to the number of loop iterations. We first assume there are no nested loops; nested cases will be addressed later in this chapter. A basic block may occupy only a small time during each loop iteration, and representing its execution time with a single interval may lead to inaccuracies, using the earliest possible start time in the first iteration and the latest possible end time in the last iteration. For example, as shown in Fig.~\ref{fig:innerLoopReprent}(a), the execution time of basic block $b_2$ may be represented as $[b2\ iter\ 1\ start,\ b2\ iter\ n\ end]$. However, this interval may include many time segments during which $b_2$ cannot actually execute—for instance, the interval $[b2\ iter\ 1\ not\ execution]$ in the first iteration. To exclude such infeasible regions and improve the precision of the representation while maintaining acceptable computational complexity, we assume that each basic block may execute in every iteration. Accordingly, we maintain a separate interval for each iteration. This is like loop unrolling, as illustrated by the transformation from Fig.~\ref{fig:innerLoopReprent}(b) to Fig~\ref{fig:innerLoopReprent}(c), where the nodes $b_2$ through $b_6$ in Fig~\ref{fig:innerLoopReprent}(c) correspond to the loop body $L_B$ with the back edge $\langle b_6, b_2 \rangle$ removed. Use $L_I^k$ to denote the immediate enclosing loop of basic block k and $N_{L_I^k} $ to denote the iteration upper bound. Therefore, the $BBOTime(k)$ is represented as a sequence of intervals, as defined in Eqn.~\ref{label:eq2}. The $i$-th element of the sequence, $BBMET(k, i)$, denotes the interval $[IBBEST(k, i),\ IBBLET(k, i)]$, which corresponds to the earliest start time and the latest end time of $k$ during the $i$-th iteration of loop $L_I^k$.
\begin{align}
    \{BBMET(k, i), ..., BBMET_N(k, i)\}, i \in [1, N_{L_I^k}]
    \label{label:eq2}
\end{align}

To simplify the representation of $BBOTime(k)$, we introduce several auxiliary definitions. Let $BBSC(x, y)$ denote the short execution cost along the control flow path from node $x$ to node $y$, excluding the execution cost of node $y$; similarly, $BBLC(x, y)$ denotes the corresponding long cost. Let $LPSC(L_Y)$ and $LPLC(L_Y)$ represent the short and long execution costs, respectively, from the loop head to the loop tail within loop $L_Y$. Let $L_Y(k)$ denote a basic block $k$ within loop $L_Y$. If $k=h$, it denotes the head node of the loop. The notation $iter_i(L_Y, k)$ refers to the $i$-th iteration instance of $L_Y(k)$. Computing $IBBEST(k, i)$ is equivalent to finding the shortest execution path from the first unrolled instance of basic block $b_2$ to the $i$-th unrolled instance of basic block $k$ in the unrolled CFG shown in Fig.~\ref{fig:innerLoopReprent}(c). This can be expressed as $BBSC(iter_1(L_Y, b_2), iter_i(L_Y, k))$. Since only one control flow exists between consecutive unrolled instances, we can first compute the local shortest path between adjacent $b_2$ nodes, then compute the shortest path from $b_2$ in the $i$-th iteration to basic block $k$ within the same iteration, and finally combine the results to obtain the global shortest path. Given that each unrolled instance is structurally identical, $IBBEST(k, i)$ can be expressed using Eqn.~\ref{label:eq3}. Similarly, $IBBLET(k, i)$ can be represented by Eqn.~\ref{label:eq4}. Consequently, the $BBOTime(k)$ can be expressed as the interval sequence shown in Eqn.~\ref{label:eq5}.
\begin{align}
    &(i - 1) * \mathtt{LPSC}(L_I^k) + \mathtt{BBSC}(L_I^k(h), k)  \label{label:eq3} \\ 
    &(i - 1) * \mathtt{LPLC}(L_I^k) + \mathtt{BBLC}(L_I^k(h), k) + \mathtt{BBWC}(k) \label{label:eq4} \\ 
    &[IBBEST(k,i), IBBLET(k, i)], i \in [1, N_{L_I^k} - 1] \label{label:eq5}
\end{align} 

\begin{figure}
    \centering
    \includegraphics[width=0.95\linewidth]{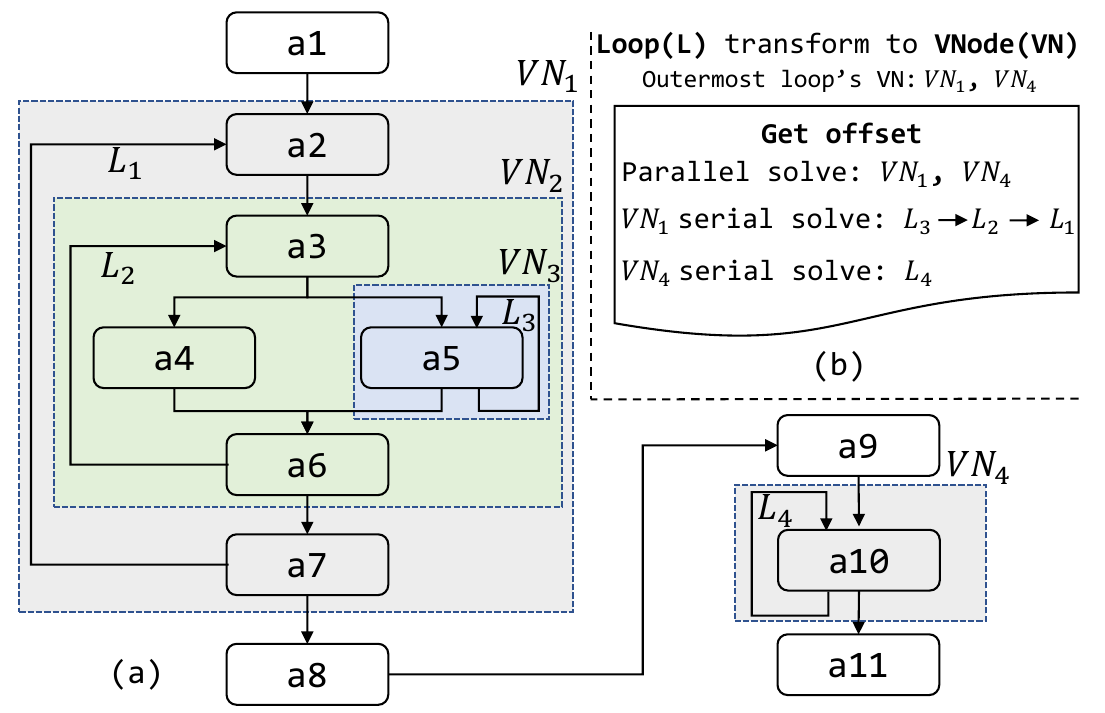}
    \caption{Loops are transformed into virtual nodes and analysis order for the entire program}
    \label{fig:compute}
    \vspace{-10pt}
\end{figure}

The outermost loops have no relative time ($LPRTime$); instead, their base time ($LPBTime$) is represented as an interval. We treat loop $k$ as a virtual node $V_k$, preserving the edges that connect it to blocks outside the loop. As illustrated in Fig.~\ref{fig:compute}, loop $L1$ is transformed into a virtual node $VNode\ 1$, abbreviated as $V_1$. The base time of $V_k$ is represented similarly to the offset time of basic blocks outside any loop, using a single interval. The lower bound of this interval is the earliest possible start time of $V_k$, denoted as $BBESOT(V_k)$, and the upper bound is the latest possible start time, denoted as $BBLSOT(V_k)$.

For inner loops, the relative time ($LPRTime$) is expressed as a sequence of intervals. Similarly, loop $k$ is represented as a virtual node $V_k$, and its relative time is denoted as $LPRTime(V_k)$. This is analogous to the offset time representation of basic blocks within loops and is described using an interval sequence. In this sequence, the $i$-th interval has a lower bound corresponding to the earliest possible start time of $V_k$ in the $i$-th iteration, denoted as $ILPEST(k,i)$, as expressed by Eqn.~\ref{label:eq6}. The upper bound corresponding to the latest possible start time, denoted as $ILPLST(k,i)$, as expressed by Eqn.~\ref{label:eq7}. $L_P^k$ denotes the immediate parent loop of $k$. For instance, in Fig.~\ref{fig:compute}, the immediate parent of loop $L_4$ is loop $L3$. $LPRTime(V_k)$ is formally defined in Eqn.~\ref{label:eq8}.
\begin{align}
    & \mathtt{LPSC}(L_P^k) * (i-1)  + \mathtt{BBSC}(L_P^k(h), V_k)\label{label:eq6} \\
    & \mathtt{LPLC}(L_P^k) * (i-1) + \mathtt{BBLC}(L_P^k(h), V_k)  \label{label:eq7} \\    
    &[ILPEST(k,i), ILPLST(k,i)], i \in [1, N_{L_P^k} - 1]  \label{label:eq8}
\end{align} 

The innermost loops must be processed first to correctly handle nested loops and ensure that inner loop information is available when analyzing outer loops. As shown in Fig.~\ref{fig:compute}, the processing order for the nested loop structure $L_1$ is $(L_3 \rightarrow L_2 \rightarrow L_1)$. Loops at the same nesting level can be processed in parallel—for example, $L_1$ and $L_4$ can be handled concurrently. For each loop, the earliest and latest possible execution times from the loop head to all other basic blocks are computed using a shortest path algorithm, resulting in values such as $BBSC()$, $BBLC()$, $LPSC()$, and $LPLC()$. During these computations, $BBBC()$ and $BBWC()$ are used as the weights of the edges for best-case and worst-case cost, respectively. After processing a loop, it is abstracted as a virtual node—e.g., loop $L_3$ is replaced by virtual node $VN_3$ in Fig.~\ref{fig:compute}. The execution cost of this virtual node is set to $LPSC() \times MinBd()$ for its $BBBC()$, and $LPLC() \times MaxBd()$ for its $BBWC()$. Nested loops are handled in reverse topological order of the loop nesting hierarchy. For instance, to process loop $L_1$, we must first process $L_3$, then $L_2$, and finally $L_1$.

Complexity Analysis. Assuming at most $V$ basic blocks and $E$ edges in a loop, each loop computes at most $2*V$ single-source shortest or longest paths. Each basic block or virtual node maintains the variables $BBSC$ and $BBLC$; each loop maintains the variable $LPSC$ and $LPLC$. So for a program with $P$ loops, the time complexity is $O(PEVlgV)$ and the space complexity is $O(PV)$.

\subsection{Overlapping Judgment}
To determine whether memory access behaviors from different cores interfere with each other, it is necessary to examine their spatial and temporal overlaps. Spatial overlap refers to whether two accesses are mapped to the same cache set, which can be efficiently determined by mapping from memory addresses to cache addresses. Therefore, this section focuses primarily on analyzing temporal overlap—whether the lifetimes of two accesses overlap in time. As previously described, the temporal characteristics of a basic block (or virtual node) relative to its immediate enclosing loop are captured by $BBOTime$ and $LPRTime$. This section presents a hierarchical temporal overlap analysis method with acceptable computational complexity, which integrates these intra-program temporal representations with the program’s absolute start time relative to the system ($PRSTime$), to assess whether interference may occur.

\begin{algorithm}[t]
    \caption{algorithm of Judge interval list overlap}
    \label{alg:AOA}
    \renewcommand{\algorithmicrequire}{\textbf{Input:}}
    \renewcommand{\algorithmicensure}{\textbf{Output:}}
    \begin{algorithmic}[1]
        \REQUIRE $A:List[(left, right)$, $B:List[(left, right)]$, 
        \ENSURE bool    
        \STATE merge(A)  \text{ // The merging of overlapping intervals in} \newline \text{ordered interval sequences has a time complexity of O(n).}
        \STATE merge(B)
        \STATE  i := 0, j := 0
         
        \REPEAT 
            \STATE lo := max(A[i].left, B[j].left)
            \STATE hi := min(A[i].right, B[j].right)
  
            \STATE \textbf{if} \textit{lo $<$ hi}  \textbf{then} 
            \STATE \textbf{\ \ return} true // Finish when find an overlap
            
            \STATE \textbf{if} \textit{A[i].right $<$ B[j].right}  \textbf{then} i++
            \STATE \textbf{else} j++
            
        \UNTIL $i < A.length \ \&\& \ j < B.length$
        \RETURN false  // No overlap 
    \end{algorithmic}
    \label{alg:alg1}
    
\end{algorithm}

The program's $PRSTime$ is equivalent to the release time of the corresponding task instance (job) within its chain and period. The computation of $PRSTime$ differs depending on whether the system follows a time-triggered or event-triggered model. For the back-to-back time-triggered chains considered in this work, the release time of the $k$-th instance of the $i$-th task in the merged chain $\Gamma_c$ is a fixed value and can be expressed as $k \times T_{\Gamma} + off_i$, where $T_{\Gamma}$ is the period of the chain and $off_i$ is the predefined offset of the task. In contrast, for event-triggered chains, the release time of the first task in the $k$-th period is given by $k \times T_{\Gamma}$. The release time of the $i$-th task in the chain depends on the response times of the preceding tasks and is represented as an interval derived from their best-case and worst-case response times. Specifically, it is given by  $[k \times T_{\Gamma} + \sum_{j = 1}^{i - 1}BCET_i, k \times T_{\Gamma} + \sum_{j = 1}^{i - 1}WCET_i]$, where $BCET_j$ and $WCET_j$ denote the best-case and worst-case execution times of the $j$-th task in the chain, respectively.

Checking for overlaps in the lifetimes of memory access behaviors equal to determine whether the intervals or interval sequences of the basic block's $BBATime$ intersect. Two intervals $t_x$ and $t_y$ are considered to overlap, denoted as $t_x \cap t_y$, if the endpoint of one interval lies within the other or if their endpoints coincide. Let $A$ and $B$ be two interval sequences. If there exist intervals $t_x \in A$ and $t_y \in B$ such that $t_x \cap t_y$, then $A$ and $B$ are considered to have overlapping execution. We define the merge of two interval sequences $A \otimes B$ as follows: $A \otimes B = \{{[a_{lb}+b_{lb}, a_{ub}+b_{ub}]|\forall[a_{lb},a_{ub}] \in A, \forall[b_{lb},b_{ub}] \in B} \}$. If $|A| = n_A$ and $|B| = n_B$, then the merged set contains $|A \otimes B| = n_A \times n_B$ intervals. To determine whether two interval sequences of sizes $m$ and $n$ potentially overlap, an $O(m+n)$ time complexity algorithm is proposed, as described in Algorithm~\ref{alg:alg1}. This algorithm first merges overlapping intervals within each sequence and then checks for the existence of overlapping pairs by comparing interval bounds.

The $BBATime(k)$ of basic block $k$ is derived according to Eqn.~\ref{eqn:eq9} and distinguishes between two cases: whether the block resides inside or outside a loop. For blocks within loops, the computation depends on $LPBTime()$, which is recursively derived from the loop’s relative time $LPRTime()$. Here, $V_{k-1}$ denotes the virtual node corresponding to the immediate parent loop of $V_k$. If $L_k$ is the outermost loop, the function $Outest(L_k)$ returns True.
\begin{flalign}
&\    BBATime(k)  =  &
\nonumber
\end{flalign}
\vspace{-0.8cm} 
\begin{flalign} 
	\begin{cases}
		\begin{aligned}
			& PRSTime \otimes [BBESOT(k) ,  BBLEOT(k)], k \notin loop\\
			& PRSTime \otimes LPBtime(V_k) \otimes BBOTime(k), k\in loop\\
		\end{aligned}
            \label{eqn:eq9}
	\end{cases}
\end{flalign}

The number of elements in $BBATime(k)$ is related to the loop nesting depth and the number of iterations. For basic blocks outside of any loop, $BBATime(k)$ contains only a single interval and can be directly used as input to Algorithm~\ref{alg:alg1}. However, for basic blocks within loops, $BBATime(k)$ may contain many intervals, making it computationally expensive to apply Algorithm~\ref{alg:alg1} directly. 
\begin{flalign}
&\    LPBTime(V_k)  =  &
\nonumber
\end{flalign}
\vspace{-0.8cm} 
\begin{flalign} 
	\begin{cases}
		\begin{aligned}
	   			&[BBESOT(V_k), BBLSOT(V_k)], Outest(L_k) = true \\
	   			& LPRTime(V_k) \otimes LPBTime(V_{k-1}) , otherwise \\
		\end{aligned}
        \label{eqn:eq10}
	\end{cases}
\end{flalign}

To address this, we adopt a hierarchical overlap detection approach for such cases.

Phase I: Overlap is first checked based on the lifetimes of the corresponding jobs. If the lifetimes of the jobs to which the memory access behaviors belong do not overlap, it can be concluded that the memory accesses do not interfere with each other. Otherwise, the analysis proceeds to Phase II.

Phase II: Overlap is assessed using the lifetime of the outermost loop to which the basic block belongs. Based on the loop nesting hierarchy, the virtual node $V_{k'}$ corresponding to the outermost loop enclosing basic block $k$ is identified. The execution interval is then approximated using the base time and worst-case execution cost of $V_{k'}$, resulting in the interval $[BBESOT(V_{k'}),\ BBLSOT(V_{k'}) + BBWC(V_{k'})]$. This interval is treated as the execution time of basic block $k$ and used as input to Algorithm~\ref{alg:alg1}. If it returns false, it indicates that the two cannot execute concurrently. Otherwise, the analysis proceeds to the next phase.

Phase III: The $BBOTime$ of the basic block is used to further determine overlaps. Since expanding $BBATime(k)$ via Eqn.~\ref{eqn:eq10} may yield many intervals, a threshold is set to ensure scalability. If the number of intervals is below the threshold, $BBATime(k)$ is used directly in Algorithm~\ref{alg:alg1}. Otherwise, an approximation is made by replacing it with the $BBATime$ of the enclosing loop’s virtual node, recursively moving outward until the interval count is acceptable. If Algorithm~\ref{alg:alg1} returns true after performing this three-step check, it indicates that their executions may overlap and are recorded in their respective overlapping arrays.

\subsection{Interference Computation}
This module computes the potential interferences $|{\mathcal M}_c(m)|$ for a memory access $m$ classified as AH or PS. First, the overlapping set of basic blocks is identified, consisting of all blocks whose memory accesses temporally overlap with $m$. Next, the cache set corresponding to $m$ is determined using the memory-to-cache address mapping, a relatively straightforward process omitted here for brevity. Finally, each memory access in the overlapping set is checked to determine whether it maps to the same cache set as $m$. If so, $|{\mathcal M}_c(m)|$ is incremented accordingly.

\subsection{Mutually Exclusive Optimization}
\begin{figure}
    \centering
    \includegraphics[width=1\linewidth]{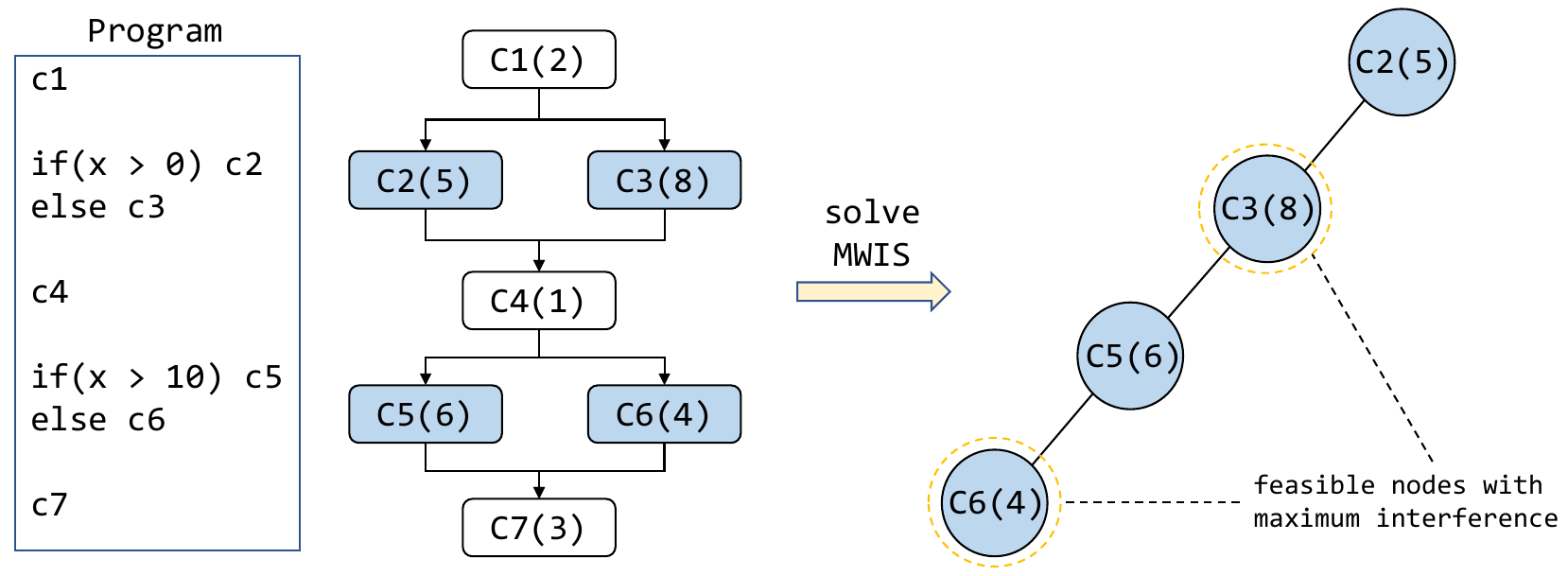}
    \caption{Transform into a Maximum Weight Independent Set problem to compute the maximum number of interferences}
    \label{fig:MWIS}
    \vspace{-10pt}
\end{figure}

Here, optimizations are performed based on mutually exclusive basic blocks from both intra-program and inter-program (chains) perspectives.

In the interference calculation described above, only the execution times of interfering basic blocks from programs running on other cores are considered, while mutually exclusive basic blocks within the interfering program are ignored. Mutually exclusive basic blocks refer to pairs of basic blocks that can never be executed concurrently within the same iteration or function call. Such pairs in Fig.~\ref{fig:MWIS} are $\langle c_2, c_3 \rangle$, $\langle c_3, c_5 \rangle$, and $\langle c_5, c_6 \rangle$. The values in parentheses indicate the number of interferences the block may cause to a certain memory access on another core. Clearly, the approach in the previous section can lead to overestimation. Take the mutually exclusive pair $\langle c_2, c_3 \rangle$ as an example. Since they share a common predecessor, their execution intervals may overlap in time, which may cause both to be considered as overlapping with the target memory access, resulting in their interference counts being summed, i.e., $5 + 8$. However, since they are mutually exclusive and cannot execute simultaneously in a single execution, the correct interference count should be the maximum of the two, i.e., $max(5, 8)$.

When multiple mutually exclusive pairs exist and are interdependent, the objective is to select a combination of basic blocks that yields the maximum cumulative interference. This problem is formulated as a maximum weight independent set problem. Based on the set of mutually exclusive pairs, a mutual exclusion execution graph $G_{EX} = (V, E)$ is constructed, where each basic block corresponds to a vertex $v \in V$, and an edge $e \in E$ is added between two vertices if the corresponding basic blocks are mutually exclusive. The goal is to find an independent set in this graph—i.e., a subset of vertices in which no two are adjacent—such that the total weight of the selected vertices is maximized. Using the example of the three mutually exclusive pairs, selecting $c_3$ and $c_6$ forms a valid independent set that results in the highest possible interference. To solve this problem, a recursive approach with memoization is employed.

Moreover, for ET chains, each subsequent task can only be triggered and executed immediately after the completion of its predecessor. Since the execution time of the predecessor task typically falls within a time interval, the actual start time of the subsequent task also becomes uncertain. As a result, the execution intervals of adjacent tasks may overlap, potentially leading to redundant interference calculations in the analysis. In reality, however, only one task can be executed on a core at any given time, and such interferences do not occur in actual systems. The following approach has been adopted to eliminate such analytical redundancy. Let $\Gamma_l$ denote the cause-effect chain on the interfering core, consisting of the task set ${\tau_1, \tau_2, \dots, \tau_n}$. For the memory access under analysis, denoted as $m$, we define $Intf(\tau_i, m)$ as the number of potential interferences caused by task $\tau_i$—that is, the total number of interferences that may be triggered by basic blocks within $\tau_i$ whose execution intervals overlap in time with that of the basic block containing $m$. The actual upper bound of interference experienced by $m$ is then given by: $Intf(m) = \max_{\tau_i \in \Gamma_l} Intf(\tau_i, m)$. This effectively avoids redundant interference counting caused by overlapping execution intervals of tasks on the same core.

\subsection{Latency Computation}
The pipeline analysis and ILP-based TSC-WCET computation adopt existing methods~\cite{li2006modeling,li2007chronos} and are not elaborated in this paper. In the following, we describe how to derive the maximum end-to-end latency. Specifically, after performing ILP-based analysis, we obtain the TSC-WCETs of all possible instances of each task on the chain within one hyperperiod. As established earlier, under the system model considered in this work, the sequence of task instances that form a chain on a given core is unique. However, chains on different cores may have different periods, so their task instance chains may experience different interferences within a hyperperiod. As a result, the TSC-WCET of these instances may vary, leading to different end-to-end latencies across instances. The maximum among them is selected as the MEL.

For an ET chain $\Gamma_{ET}$, let $NSI(\Gamma_{ET})$ denote the number of scheduling instances within the hyperperiod. The MEL of $\Gamma_{ET}$ can be reformulated as follows: among all instances, the one with the largest sum of the TSC-WCETs of all tasks on $\Gamma_{ET}$, which is given by:
\[
\max_{1 \leq k \leq NSI(\Gamma_{ET})} 
    \sum_{\tau_i \in \Gamma_{ET}} TSC\text{-}WCET_k(\tau_i)
\]

For a TT chain $\Gamma_{TT}$, let the last node on the chain correspond to the task $tail(\Gamma_{TT})$. As described earlier, the release offset of a task $\tau$ is denoted by $off(\tau)$. Then, the MEL of the chain $\Gamma_{TT}$ is given by:
\[
\max_{1 \leq k \leq NSI(\Gamma_{TT})} 
\left\{
    off(tail(\Gamma_{TT})) +  TSC\text{-}WCET_k(\Gamma_{TT})
\right\}
\]

\section{Proof of Safety}
The pipeline analysis, ILP formulation, and flow analysis adopted in this work are based on existing approaches whose soundness has been formally established in~\cite{li2006modeling, theiling1998combining,gustafsson2006automatic}. Therefore, we focus on the correctness of our shared cache analysis, specifically proving that the estimated number of interferences for any memory access is no less than that encountered in any possible concrete execution. In the following, we demonstrate that our method satisfies this safety property.

We begin by proving that the inter-core context constructed in our analysis is safe, i.e., the estimated memory access time intervals always cover all possible access times that may occur in actual execution. It is sound to approximate the time when a memory access occurs during a particular execution of a basic block by the possible execution window of that block instance. Furthermore, using a basic block's earliest start time and latest end time as its lifetime safely over-approximates its execution interval, accounting for variations caused by different control-flow paths. During initialization, the pipeline analysis adopts conservative assumptions—assuming all accesses to shared cache result in misses—ensuring that the estimated execution time is no less than the actual. Similarly, the best-case execution is underestimated. Therefore, our approach safely captures all possible execution intervals of a basic block, guaranteeing that the constructed context is conservative and safe.

Next, we prove that the estimated number of interferences is no less than the actual number. In concrete execution, a memory interference can only occur if the interfering access occurs strictly between the current access and its immediately preceding access to the same cache block. This forms a necessary condition for real interferences. However, our method adopts a more conservative rule: an interference is assumed whenever the lifetimes of two memory accesses to the same cache block overlap, regardless of whether they fall precisely within the narrow interference window of actual execution. As a result, the estimated number of interferences includes all possible real interferences, and potentially more. Hence, the interference model is a sound over-approximation.

Therefore, the shared cache analysis is safe, ensuring that the resulting TSC-WCET is also safe, which in turn guarantees the safety of the end-to-end latency estimation.

\section{EVALUATION}

\subsection{Setup}

\begin{figure}
    \centering
    \includegraphics[width=1\linewidth]{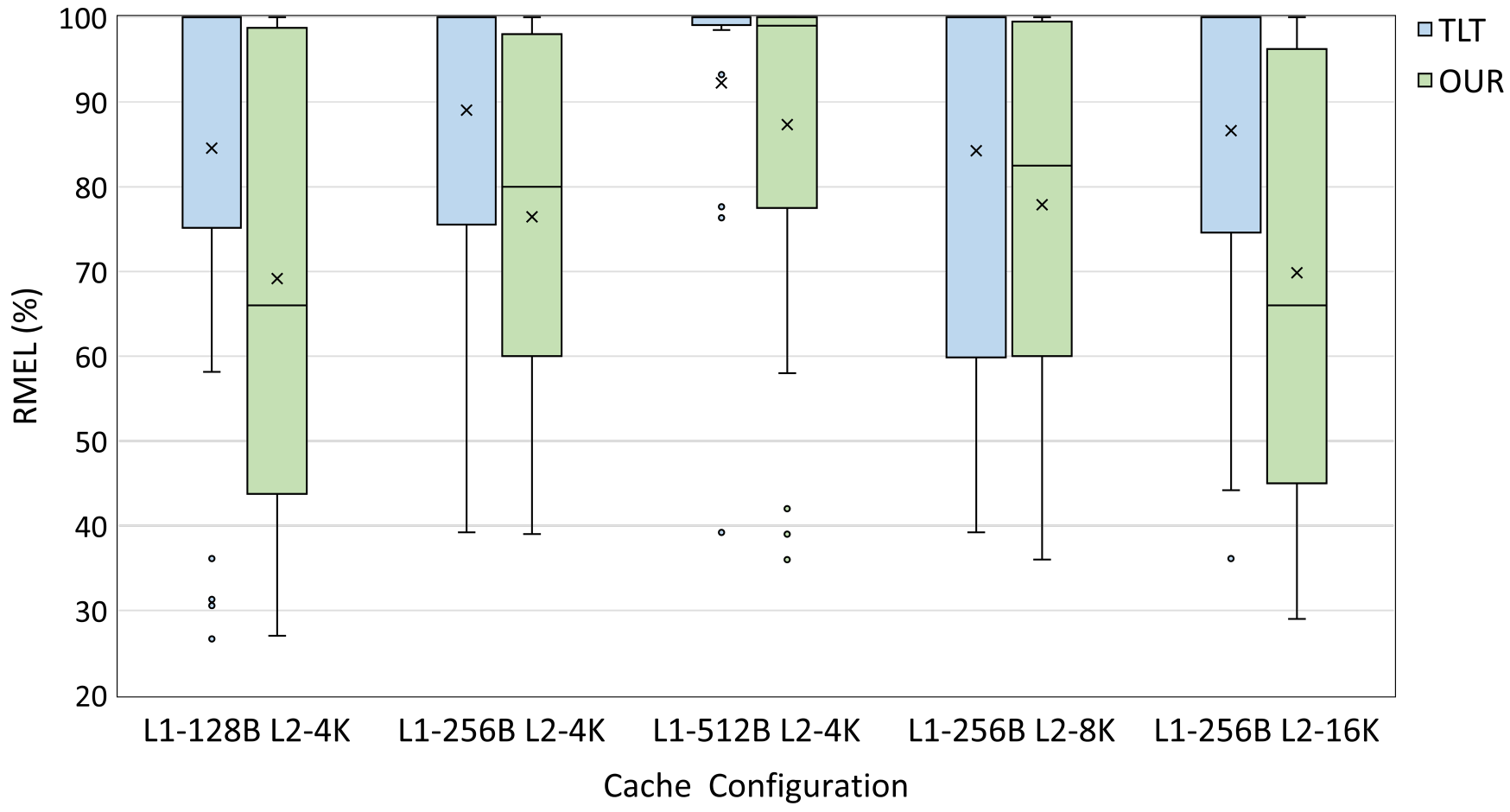}
    \caption{RMTL for one task's  Chains on Dual-core Systems with Utilization 0.9 and Various Cache Configuration
    }
    \label{fig:exp1}
    \vspace{-10pt}
\end{figure}

\begin{figure}
    \centering
    \includegraphics[width=1\linewidth]{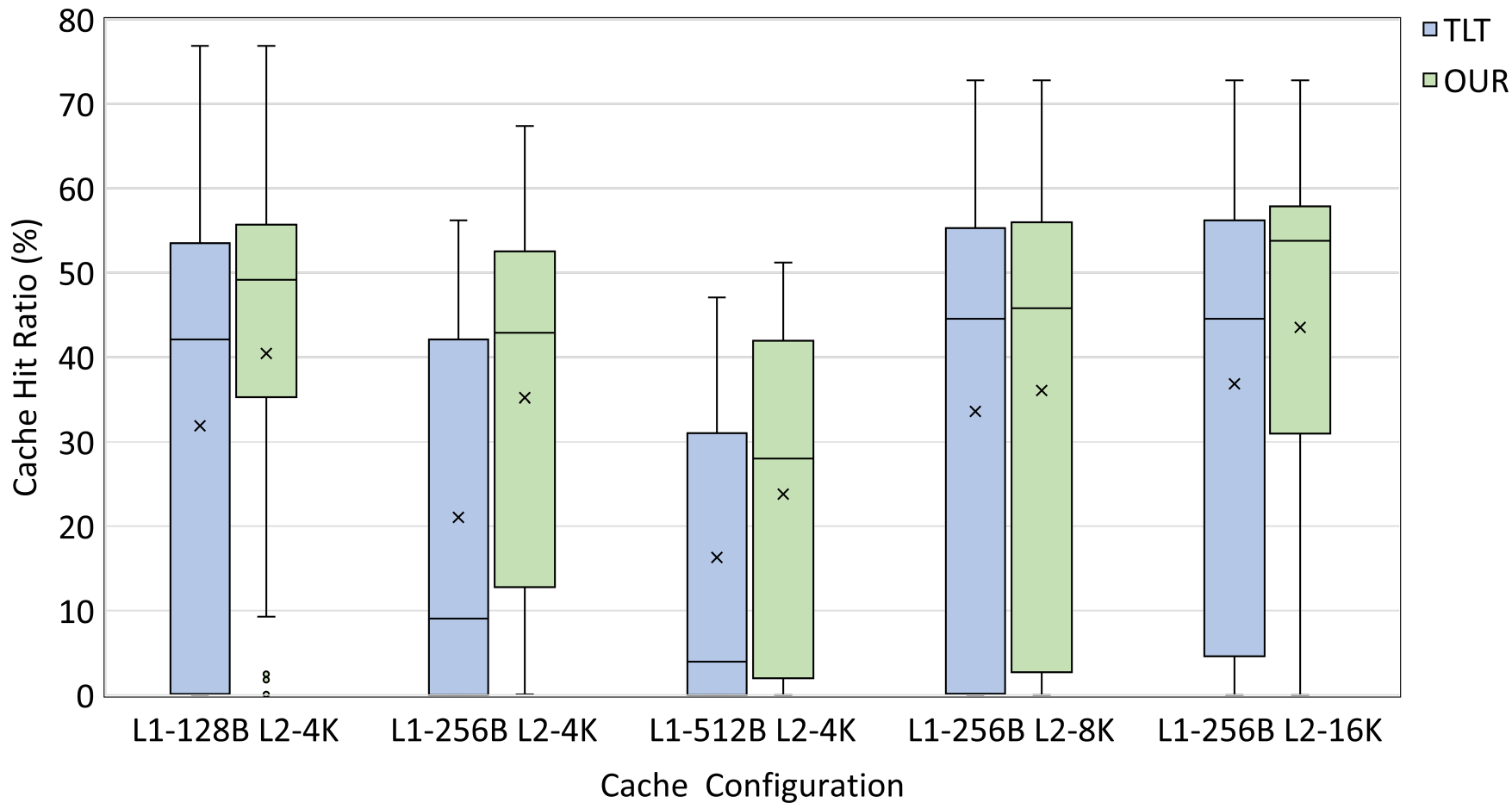}
    \caption{Cache Hit Ratio for one task's Chains on Dual-core Systems with Utilization 0.9 and Various Cache Configuration}
    \label{fig:exp2}
    \vspace{-10pt}
\end{figure}

To evaluate the effectiveness of the proposed approach, we select two representative existing methods as baselines: the No Context (NCT) method and the Task Lifetime (TLT) method~\cite{liang2012timing}. The NCT does not consider any execution context information. To ensure the safety of the analysis results, it operates under the most conservative assumption—that all accesses to the shared cache are treated as misses (assuming no timing anomalies exist in the hardware, as is the case in this work). In contrast, the TLT determines interference based on task lifetime information: if the lifetimes of two tasks overlap in time and their memory accesses map to the same cache set across different cores, they are considered to interfere. To improve the precision of this model, we extend TLT to capture task-instance-level (job-level) information, enabling a more fine-grained modeling of interference relationships. 

We build the end-to-end latency analysis framework presented in this work on top of two existing WCET analysis frameworks: Chronos~\cite{li2007chronos} and SWEET~\cite{lisper2014sweet}. Chronos primarily focuses on microarchitectural-level modeling and analysis, for which we customize its modules, including pipeline analysis, private cache analysis, and ILP-based WCET computation. The SWEET framework specializes in program-level flow analysis, where we leverage it to extract flow facts such as loop bounds and mutually exclusive basic blocks. We manually annotate cases where SWEET fails to infer loop bounds. The remaining modules in the analysis framework, such as getting context, overlapping judgment, latency calculations, etc., are implemented independently by us.

We conduct our analysis on dual-core and quad-core systems. Each core is assumed to have a simple in-order pipeline and a private L1 cache, while all cores share a L2 cache. Both levels of cache are configured as set-associative with 4 ways and employ the LRU replacement policy. Regarding access latency, we assume an L1 cache hit latency of 1 cycle, an L2 cache hit latency of 6 cycles, and an L2 cache miss latency of 30 cycles. To comprehensively evaluate the proposed method, we consider caches with varying capacities in our experiments. Detailed parameters are provided later.

Task Set Construction. The chains are generated using independent tasks from a task set~\cite{gustafsson2010malardalen}, and dependency relations are added to form cause-effect chains, following an approach similar to that in~\cite{kohler2023robust}. Each chain consists of 1, 2, or 4 tasks. To ensure the schedulability of ET chains, we adopt a method that combines context-independent pessimistic CIP-WCET with a set of predefined automotive periods $\mathcal{T}{\text{auto}}$ from automotive system benchmarks~\cite{Autobenchmarks}. First, the total CIP-WCET of all tasks in an ET chain $\Gamma_{ET}$ is computed as $T_w = \sum_{\tau_i \in \Gamma_{ET}} \text{CIP-WCET}(\tau_i)$, where CIP-WCET is derived under the assumption that all accesses to the shared cache result in misses, thereby ensuring safety and context-independence. Then, the actual triggering period $T$ is selected as the smallest value in $\mathcal{T}_{\text{auto}}$ that is no less than $T_w$, i.e., $T = \min \left\{ T' \in \mathcal{T}_{\text{auto}} \mid T' \geq T_w \right\}$. For TT chains $\Gamma_{TT}$ with a back-to-back characteristics, the period and task offset times are also determined based on CIP-WCET. Specifically, the offset time $off(i)$ of the $i$-th task is set to the cumulative sum of the CIP-WCETs of the preceding $i-1$ tasks to enforce back-to-back execution semantics: $off(i) = \sum_{j=1}^{i-1} \text{CIP-WCET}(\tau_j)$. The period for the TT chain is then obtained using the same method as the ET chains.

We adopts the Relative MEL (RMEL) as the primary evaluation metric, where the MEL obtained under the NCT method is used as the baseline, and the RMEL is computed as the ratio between the evaluated method and this baseline. A smaller RMEL value indicates a more significant improvement. In addition, the shared cache hit rate is introduced as a complementary metric to provide a more comprehensive assessment of the effectiveness of cache modeling optimizations. Compared to RMEL, which only reflects cache modeling improvements along the critical path that determines task WCET, the cache hit rate can better capture optimizations on non-critical paths that may be overlooked.

\subsection{End to End Latency Evaluation}
We evaluate dual-core and quad-core systems under various cache and task chain configurations. For each task count configuration, 40 task sets are randomly generated as analysis samples. Experimental results demonstrate that, across all tested configurations, our method consistently outperforms the TLT approach regarding both RMEL and cache hit rate. Specifically, under certain configurations, the proposed method achieves up to a 66\% improvement in RMEL on dual-core systems and 74\%on quad-core systems.

\subsubsection{dual‑core systems}
\
\newline
\indent
We first analyze the performance of our method on dual-core systems under varying task counts and core utilization.

\textbf{Single-task case.} In this scenario, TT and eET chains are essentially equivalent, as the task in both types of chains is periodically triggered. We analyze the performance of different methods under varying cache configurations and utilization. The cache configurations are shown in Fig.~\ref{fig:exp1}, which also presents the RMEL at a core utilization of 0.9. As observed from the boxplots, our proposed method consistently outperforms. In each box, the horizontal line indicates the mean, while the cross represents the median. For instance, under the configuration with an L1 cache of 256B and an L2 cache of 4KB, the average RMEL of the TLT method is 89\%, whereas our method achieves 76\%. Fig.~\ref{fig:exp2} also shows that our method significantly improves the shared cache hit rate; under the same configuration, the average hit rate of our method is 35\% compared to 26\% for TLT. When only reducing the size of the L1 cache, both TLT and our method show a decrease in RMEL, with our method maintaining a clear advantage. This is supported by Fig.~\ref{fig:exp2}, which shows an increase in cache hit rate as the L1 cache size decreases.

\begin{figure}
    \centering
    \includegraphics[width=1\linewidth]{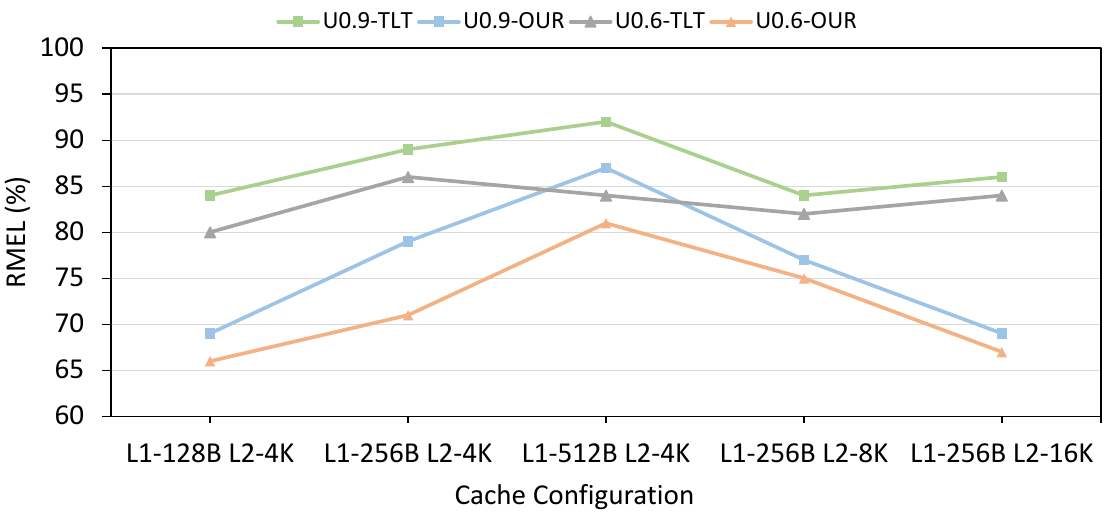}
    \vspace{-20pt}
    \caption{RMTL for one task's Chains on Dual-core Systems with Utilization 0.6 and 0.9 and Various Cache Configuration}
    \label{fig:exp3}
    \vspace{-5pt}
\end{figure}

The TLT method tends to overestimate interference by summing the effects of all overlapping task instances whenever their lifetimes overlapping, even if the overlap is minimal. When the core utilization is reduced to 0.5, the two methods show a decrease in RMEL, as shown in Fig. ~\ref{fig:exp3}, and the cache hit rate improves (not shown here for brevity). Our method still yields better results. This is because lower utilization reduces task execution within a given period, thereby reducing inter-core interference during the same time window.

\begin{table}[t]
\centering
\caption{Comparison of Cache Hit Ratio for two tasks TT and ET Chains under Different Cache Configurations}
\begin{tabular}{l cc cc cc}
\toprule
\begin{tabular}[c]{@{}c@{}}Trigger\\Model\end{tabular}
& \multicolumn{2}{c}{\begin{tabular}[c]{@{}c@{}}L1-128B\\L2-4K\end{tabular}}
& \multicolumn{2}{c}{\begin{tabular}[c]{@{}c@{}}L1-256B\\L2-4K\end{tabular}}
& \multicolumn{2}{c}{\begin{tabular}[c]{@{}c@{}}L1-512B\\L2-8K\end{tabular}} \\
\cmidrule(lr){2-3} \cmidrule(lr){4-5} \cmidrule(lr){6-7}
& TLT & OUR & TLT & OUR & TLT & OUR \\
\midrule
EE & 21 & 30 & 13 & 25 & 7 & 16 \\
TT & 29 & 38 & 22 & 33 & 14 & 21 \\
\bottomrule
\end{tabular}
\label{table:exp1}
\vspace{-10pt}
\end{table}

\begin{figure}
    \centering
    \includegraphics[width=1\linewidth]{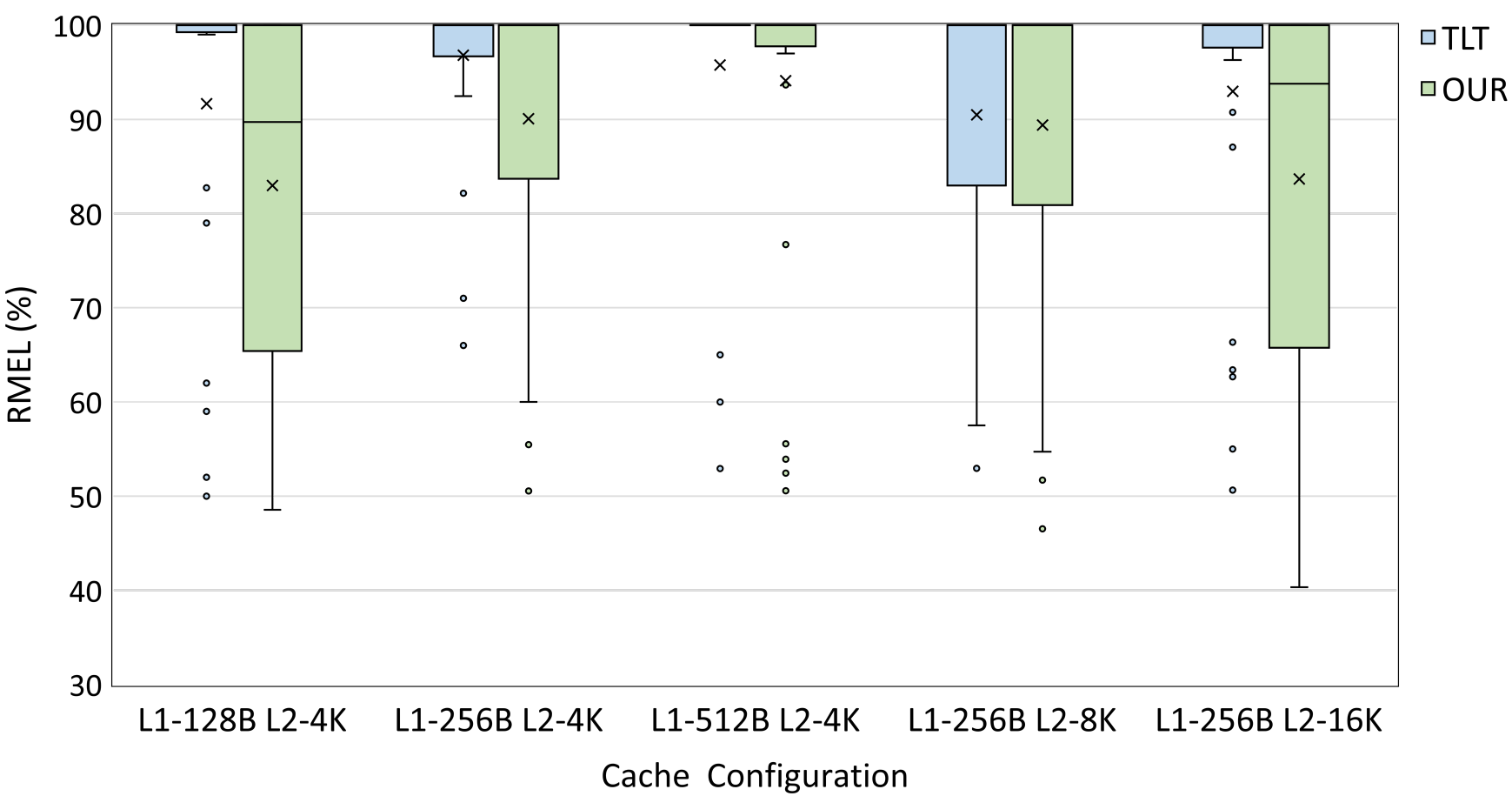}
    \caption{RMTL for two tasks TT Chains on Dual-core Systems with Utilization 0.9 and Various Cache Configuration}
    \label{fig:exp4}
\end{figure}

\begin{figure}
    \centering
    \includegraphics[width=1\linewidth]{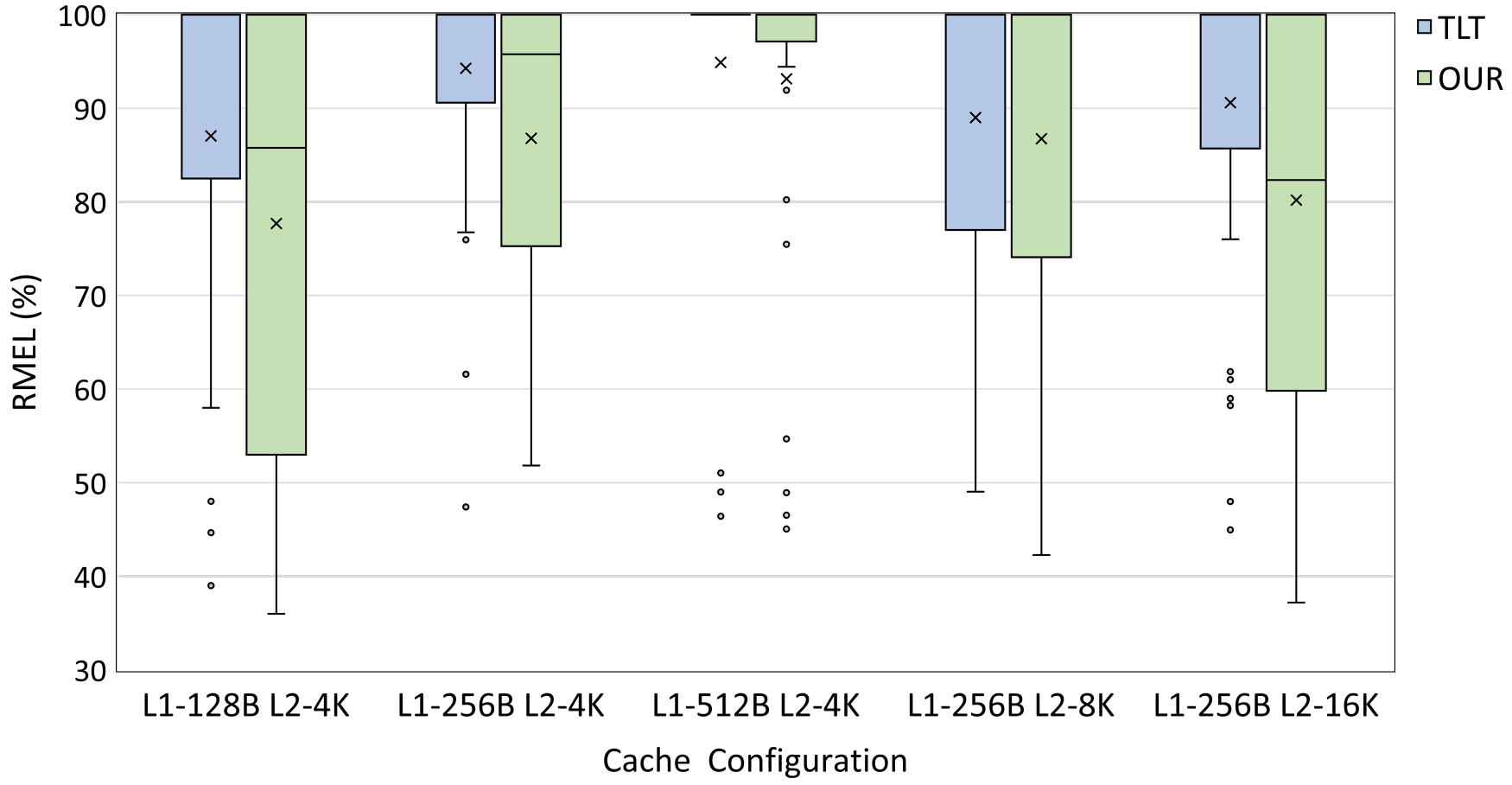}
    \caption{RMTL for two tasks ET Chains on Dual-core Systems with Utilization 0.9 and Various Cache Configuration}
    \label{fig:exp5}
    \vspace{-10pt}
\end{figure}

\textbf{Two-tasks case.} By comparing Fig.~\ref{fig:exp4} and Fig.~\ref{fig:exp5}, we observe that, in terms of RMEL, both the TLT method and our method perform relatively better on ET chains than on TT chains. In both cases, our method consistently yields better results. From the perspective of interference analysis alone, one would expect TT chains to achieve better results, since the start times of all tasks are statically defined, enabling more precise interference modeling. In contrast, for ET chains, only the start time of the first task is fixed, while the execution start times of subsequent task instances become uncertain. This can lead to overlapping execution windows among different task instances within the same chain. As a result, the TLT method may overestimate interference by accumulating contributions from multiple overlapping instances, even though they cannot execute in parallel on the same core.

The underlying reason for this counterintuitive result is that, for TT chains with fixed task offsets, the optimization of end-to-end latency is primarily reflected in the execution time of the last task in the chain. Consequently, the overall benefit is diluted across the entire chain's end-to-end latency. As shown in Table~\ref{table:exp1}, the cache hit rates for TT chains are generally higher than those for ET chains, indicating that interference in TT chains is more analyzable and predictable.

\begin{figure}
    \centering
    \includegraphics[width=1\linewidth]{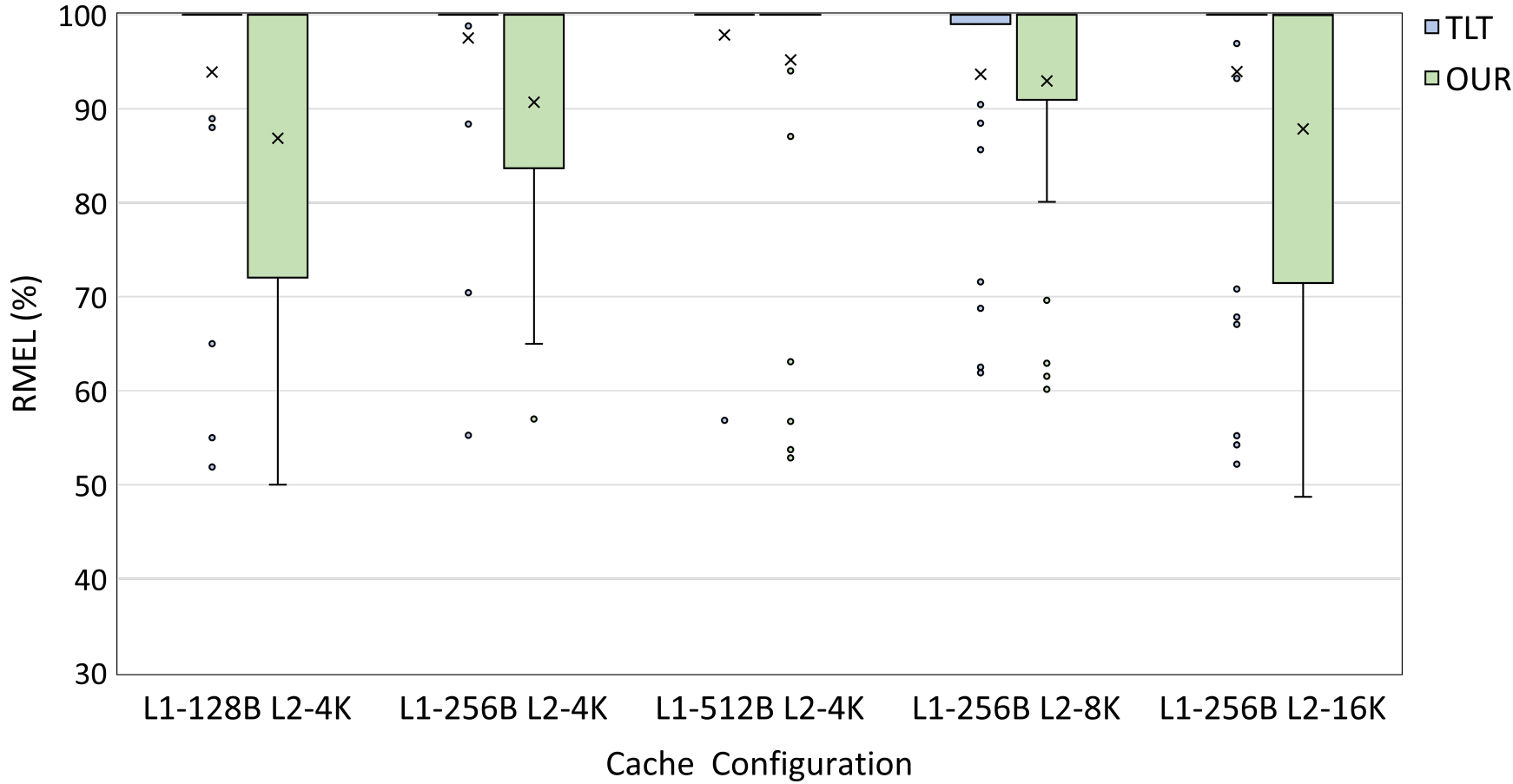}
    \caption{RMTL for four tasks ET and TT Chains on Dual-core Systems with Utilization 0.9 and Various Cache Configuration}
    \label{fig:exp6}
\end{figure}

\textbf{Four-task case.} To present the analysis results concisely, we combine the results for both TT and ET chains in Fig.~\ref{fig:exp6}. It can be observed that, compared to the one-task and two-task chains, the optimization gains in end-to-end latency for both methods decrease as the number of tasks in the chain increases. This reduction is primarily due to two factors. First, as the number of tasks grows, the TLT method becomes increasingly pessimistic in its interference analysis for ET chains, thereby reducing its analytical precision. Second, for TT chains, the increased number of tasks tends to dilute the impact of final-task execution time optimization on the overall end-to-end latency, resulting in a diminished overall improvement. Nevertheless, the experimental results consistently demonstrate that our method outperforms the TLT method across all configurations.

\begin{table*}[htbp]
\centering
\caption{RMTL and Cache Hit Ratio for ET and TT Chains on Quad-core Systems with Utilization 0.9 and Various Configuration}
\begin{tabular}{cc cccc cccc cccc}
\toprule
\multirow{3}{*}{\textbf{Task Number}} & \multirow{3}{*}{\textbf{Metrics}}
    & \multicolumn{4}{c}{\textbf{L1-256B L2-4K}}
    & \multicolumn{4}{c}{\textbf{L1-256B L2-8K}}
    & \multicolumn{4}{c}{\textbf{L1-256B L2-16K}} \\
\cmidrule(lr){3-6} \cmidrule(lr){7-10} \cmidrule(lr){11-14}
 & & \multicolumn{2}{c}{Best} & \multicolumn{2}{c}{Average}
   & \multicolumn{2}{c}{Best} & \multicolumn{2}{c}{Average}
   & \multicolumn{2}{c}{Best} & \multicolumn{2}{c}{Average} \\
 & & TLT & OUR & TLT & OUR
   & TLT & OUR & TLT & OUR
   & TLT & OUR & TLT & OUR \\
\midrule
1 & \multirow{3}{*}{RMEL}
  & 36 & 38 & 91 & 86
  & 37 & 36 & 88 & 74
  & 36 & 36 & 84 & 78 \\
2 &
  & 48 & 43 & 95 & 90
  & 44 & 42 & 97 & 93
  & 44 & 42 & 90 & 86 \\
4 &
  & 60 & 54 & 98 & 94
  & 57 & 52 & 96 & 93
  & 53 & 50 & 93 & 90 \\
\midrule
1 & \multirow{3}{*}{\shortstack{Cache Hit \\ Ratio}}
  & 67 & 67 & 11 & 24
  & 72 & 72 & 23 & 33
  & 72 & 72 & 33 & 37 \\
2 &
  & 54 & 65 & 8 & 21
  & 60 & 74 & 15 & 30
  & 63 & 70 & 30 & 34 \\
4 &
  & 45 & 60 & 7 & 18
  & 49 & 68 & 11 & 30
  & 55 & 70 & 25 & 30 \\
\bottomrule
\end{tabular}
\label{table:exp2}
\end{table*}

\subsubsection{quad‑core systems}
\
\newline
\indent
To evaluate the scalability of our method, we further extend the experimental platform to a quad-core. The results are presented in Table~\ref{table:exp2}. Compared to the dual-core platform, both methods exhibit increased MEL on the quad-core system, reflected by a higher upper value on RMEL. This is primarily attributed to the rise in the number of interfering cores—from one to three—which significantly intensifies the level of interference and thereby affects the cache analysis results.

Despite this, the experimental results demonstrate that our method consistently outperforms the TLT method in the quad-core scenario, confirming its advantage in terms of scalability to many-core platforms. Additionally, it is observed that the shared cache hit rates on the quad-core system are generally lower than those on the dual-core system.

It is also worth noting that as the base level of interference increases, our method—despite offering more accurate interference estimations—may still fail to fully prevent cache evictions in certain cases. For instance, if the TLT method estimates 10 interfering accesses and our method reduces this number to 6, the 4-interference reduction may still be insufficient to avoid eviction when the cache associativity is 4. Nevertheless, our method still achieves a higher cache hit rate than the TLT method.

\begin{table}[htbp]
\centering
\vspace{-10pt}
\caption{RunTime for Two Method in Minutes}
\begin{tabular}{c cc cc cc}
\toprule
Core Num & \multicolumn{2}{c}{1 Task} & \multicolumn{2}{c}{2 Tasks} & \multicolumn{2}{c}{4 Tasks} \\
\cmidrule(r){2-3} \cmidrule(r){4-5} \cmidrule(r){6-7}
         & TLT & OUR & TLT & OUR & TLT & OUR \\
\midrule
2 & 0.4 & 2.6 & 0.7 & 5.3 & 1.5 & 13.1 \\
4 & 0.9 & 6.5 & 2.2 & 10.4 & 3.2 & 18.9 \\
\bottomrule
\end{tabular}
\label{table:exp4}
\end{table}

\subsection{RunTime Evaluation}
This section further evaluates the average runtime of the TLT method and our method under varying task counts and core configurations. The results are shown in Table~\ref{table:exp4}. It can be observed that, for both methods, the analysis time increases with the number of tasks and cores, reflecting the sensitivity of analysis complexity to system scale.

Regarding overall comparison, the TLT method consistently demonstrates shorter analysis times, indicating lower computational overhead. In contrast, our proposed method incurs higher analysis times due to incorporating more fine-grained context modeling and interference evaluation mechanisms. This is particularly evident in configurations with a large number of tasks and cores—for example, in the 4-task, 4-core scenario, the average analysis time of our method reaches 18.9 minutes—highlighting the increased computational cost introduced by enhanced analysis precision.

\section{Conclusion \& Future Work}
This paper presents a novel end-to-end latency analysis framework for multi-chain systems deployed on multi-core platforms with shared caches by leveraging cause-effect chains' scheduling information and structural characteristics. The experiment results show that, compared to conventional methods, our approach significantly improves both shared cache hit rates and the accuracy of end-to-end latency.

It is worth noting that, to ensure relatively deterministic task execution sequences on each core—an essential prerequisite for shared cache analysis—certain constraints have been imposed on the scheduling model. Specifically, tasks in a chain have a unified priority, and each core is limited to chains with a single triggering mode. Additionally, multiple chains on the same core must satisfy conditions that allow them to be transformed into a single chain. To the best of our knowledge, this is the first work that integrates the structural characteristics of cause-effect chains to enable scheduling aware WCET analysis and subsequent end-to-end latency estimation. In future work, we aim to relax these assumptions and further extend the analysis framework to support more general scheduling models and more complex execution scenarios.


\bibliographystyle{IEEEtran}
\bibliography{./reference}

\section{Biography Section}

\vspace{-20pt}
\begin{IEEEbiography}[{\raisebox{0.25\height}{\includegraphics[width=1in,height=1.25in,clip,keepaspectratio]{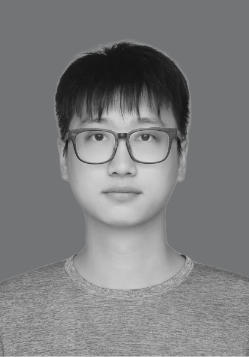}}}]{Yixuan Zhu}
received the B.S. degree in software engineering from University of Electronic Science and Technology of China, ChengDu, China, in 2022. He is currently pursuing the Ph.D degree in computer science at the University of Science and Technology of China, Hefei, China. His research interests include time behavior analysis, microarchitecture modeling, and timing anomalies.
\end{IEEEbiography}

\vspace{-20pt}

\begin{IEEEbiography}[{\raisebox{0.25\height}{\includegraphics[width=1in,height=1.25in,clip,keepaspectratio]{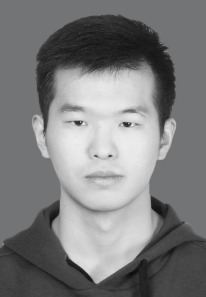}}}]{Yinkang Gao}received the B.S. degree in computer science from University of Nanjing University of Posts and Telecommunications, Nanjing, China, in 2020. He is currently pursuing the Ph.D degree in computer science at the University of Science and Technology
 of China, Hefei, China. His research interests include compueter architectures and embedded systems.
\end{IEEEbiography}

\vspace{-20pt}

\begin{IEEEbiography}[{\raisebox{0.25\height}{\includegraphics[width=1in,height=1.25in,clip,keepaspectratio]{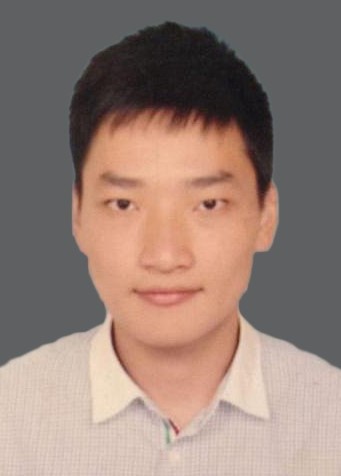}}}]{Bo Zhang}
is a Ph.D. candidate in Computer Science at the University of Science and Technology of China (USTC), Hefei, China, specializing in the analysis and optimization of real-time embedded systems. His research expertise centers on temporal behavior modeling for complex embedded architectures and cause-effect chain optimization in safety-critical
systems.
\end{IEEEbiography}

\vspace{-20pt}

\begin{IEEEbiography}[{\raisebox{0.25\height}{\includegraphics[width=1in,height=1.25in,clip,keepaspectratio]{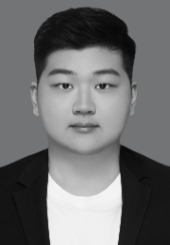}}}]{Xiaohang Gong}
eceived the B.E. degree in computer science and technology in 2022 and the M.E. degree in computer science and technology in 2025 from the University of Science and Technology of China (USTC), Hefei, China.  His research interests include real-time systems, time-predictable cache coherent systems, and memory system design.
\end{IEEEbiography}

\vspace{-20pt}

\begin{IEEEbiography}[{\raisebox{0.25\height}{\includegraphics[width=1in,height=1.25in,clip,keepaspectratio]{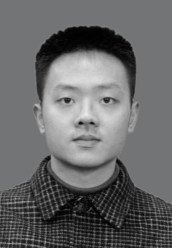}}}]{Binze Jiang}
received his master degree in computer science and technology from University of Science and Technology of China, China, in 2025. His research interests include computer architecture and real-time systems.
\end{IEEEbiography}

\vspace{-20pt}

\begin{IEEEbiography}[{\raisebox{0.0\height}{\includegraphics[width=1in,height=1.25in,clip,keepaspectratio]{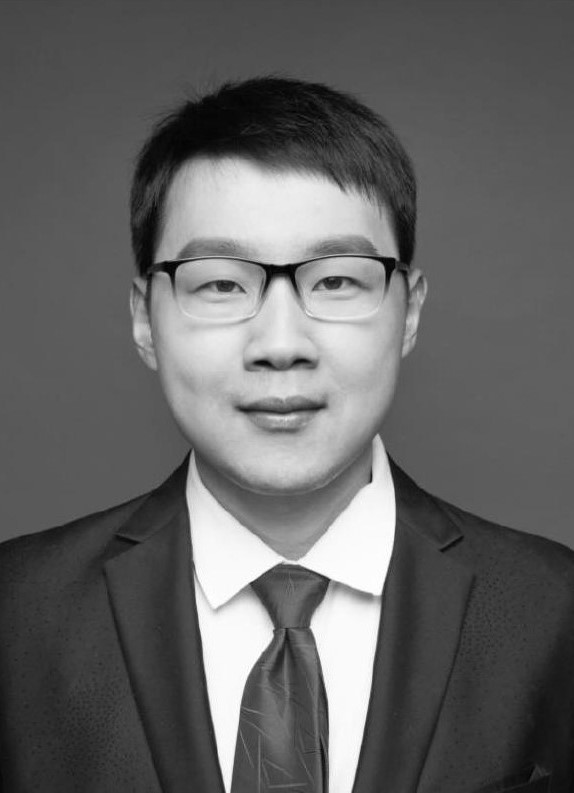}}}]{Lei Gong}
received the Ph.D. degree in computer science from the University of Science and Technology of China, Hefei, China, in 2019. He is an Associate Professor with School of Computer Science, University of Science and Technology of China. His research interests include FPGA-based accelerator designs and artificial intelligence and machine learning systems. Dr. Gong’s paper has once been nominated as Best Paper Candidate in CODES+ISSS 2018.
\end{IEEEbiography}

\vspace{-20pt}

\begin{IEEEbiography}[{\raisebox{0.0\height}{\includegraphics[width=1in,height=1.25in,clip,keepaspectratio]{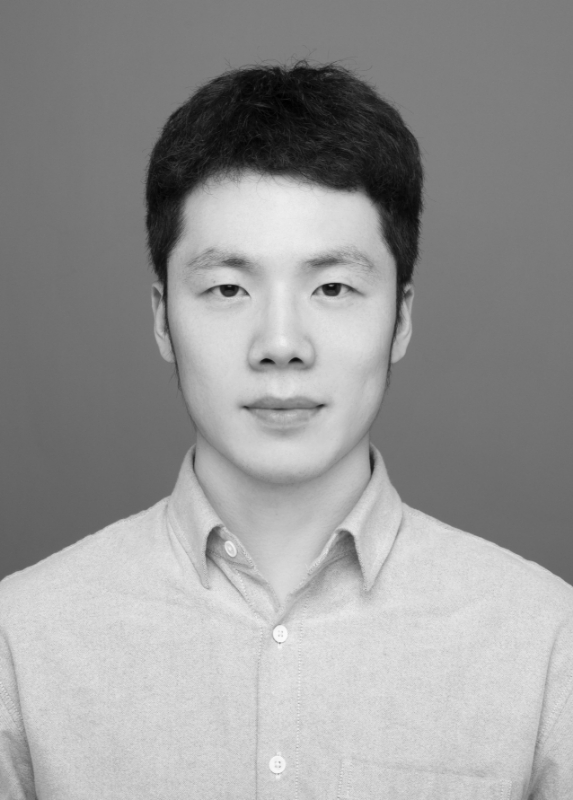}}}]{Wenqi Lou}
received the B.S. degree from Northwestern Polytechnical University, Xi’an, China, in 2018, and the Ph.D. degree in computer science from the University of Science and Technology of China, Hefei, China, in 2023. He is currently an Associate Researcher with the School of Software Engineering, University of Science and Technology of China. His current research interests include deep learning accelerators and FPGA-based acceleration.
\end{IEEEbiography}

\vspace{-5pt}

\begin{IEEEbiography}[{\raisebox{0.2\height}{\includegraphics[width=1in,height=1.25in,clip,keepaspectratio]{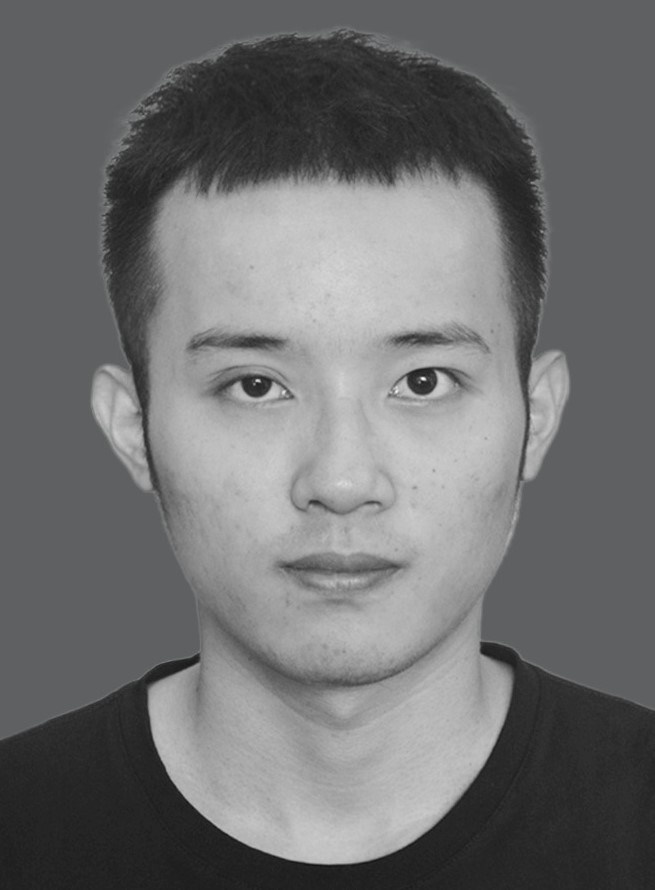}}}]{Teng Wang}
received the Ph.D. degree from the University of Science and Technology of China, Hefei, China, in 2023. He is currently a Research Scientist with Suzhou Institute for Advanced Research, University of Science and Technology of China, Suzhou, China. His research interests include algorithm-level and architecture-level optimizations of FPGA for deep learning applications.
\end{IEEEbiography}

\vspace{-15pt}

\begin{IEEEbiography}[{\raisebox{0.0\height}{\includegraphics[width=1in,height=1.25in,clip,keepaspectratio]{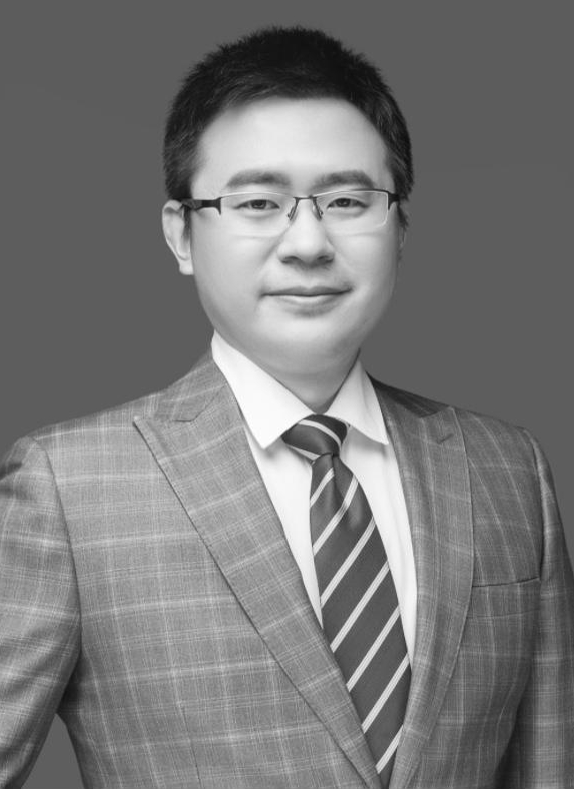}}}]{Chao Wang}
(Senior Member, IEEE) received the B.S. and Ph.D. degrees in computer science from the University of Science and Technology of China, Hefei, China, in 2006 and 2011, respectively. He is currently a Professor with the University of Science and Technology of China. His research interests include multicore and reconfigurable computing. Prof. Wang serves as the Associate Editor for ACM Transactions on Design Automation for Electronics Systems and IEEE/ACM Transactions on Computational Biology and Bioinformatics.
\end{IEEEbiography}

\vspace{-15pt}

\begin{IEEEbiography}[{\raisebox{0.0\height}{\includegraphics[width=1in,height=1.25in,clip,keepaspectratio]{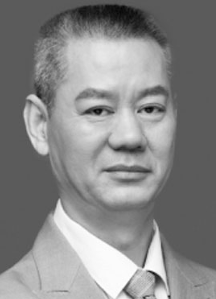}}}]{Xi Li}
received the PhD degree in computer science from the University of Science and Technology of China, in 2003, and is currently a professor in the School of Computer Science and Technology, and the School the Software Engineering, University of Science and Technology of China. There he directs the research programs in High Energy-efficiency Intelligent ComputingLab, examining various aspects of computer systems, especially real-time embedded systems.
\end{IEEEbiography}

\vspace{-15pt}

\begin{IEEEbiography}[{\raisebox{0.0\height}{\includegraphics[width=1in,height=1.25in,clip,keepaspectratio]{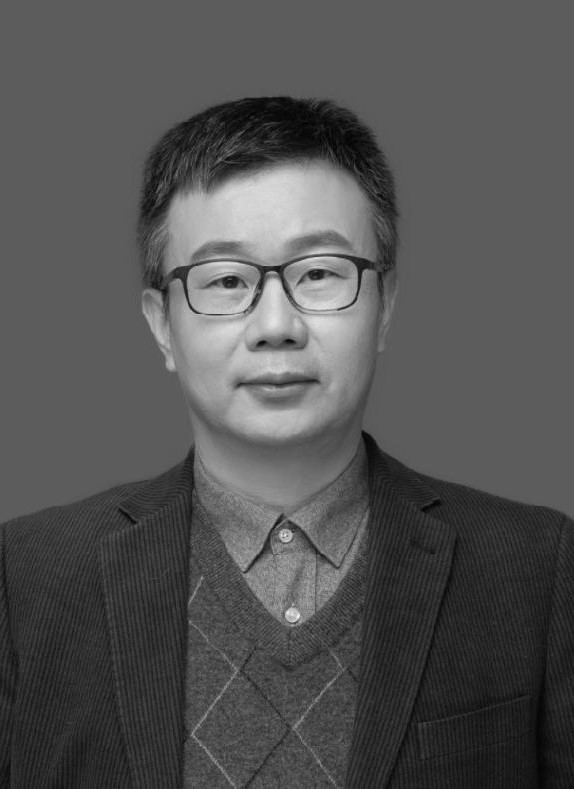}}}]{Xuehai Zhou}
received the B.S., M.S., and Ph.D. degrees from the University of Science and Technology of China, Hefei, China, in 1987, 1990, and 1997, respectively. He is currently a Professor with the School of Computer Science, University of Science and Technology of China. He serves as a General Secretary of the Steering Committee of Computer College Fundamental Lessons, and the Technical Committee of Open Systems, China Computer Federation.
\end{IEEEbiography}

\vfill

\end{document}